\documentstyle[epsfig]{basi}
\begin{document}
\title[NIR Stellar Spectral Library]{A Near-Infrared Stellar Spectral Library: II. K-Band Spectra}
\author[Arvind C. Ranade et al.]%
       {Arvind C. Ranade$^1$, Harinder P. Singh$^2$,
       Ranjan Gupta$^3$\thanks{e-mail:rag@iucaa.ernet.in}, and
       N M Ashok$^4$\\
 $^1$Vigyan Prasar, A-50, Institutional Area, Sector-62, NOIDA 201 307, India\\
 $^2$ Department of Physics \& Astrophysics, University of Delhi, Delhi 110 007, India\\
        $^3$IUCAA, Post Bag 4,Ganeshkhind, Pune 411 007, India\\
        $^4$Physical Research Laboratory, Navrangpura, Ahmedabad 380009,
        India}
\maketitle
\label{firstpage}
\begin{abstract}This paper is the second in the series of papers
on near-infrared (NIR) stellar spectral library produced by reducing
the observations carried out with 1.2 meter Gurushikhar Infrared
Telescope (GIRT), at Mt. Abu, India using a NICMOS3 HgCdTe $256
\times 256$ NIR array based spectrometer. In paper I (Ranade et al.
2004), H-band spectra of 135 stars at a resolution of $\sim 16$\AA~
were presented. The K-band library being released now consists of
114 stars covering spectral types O7--M7 and luminosity
classes I--V. The spectra have a moderate resolution of $\sim
22$\AA~ in the K band and have been continuum shape corrected to
their respective effective temperatures. We hope to release the
remaining J-band spectra soon. The complete H and K-Band library is
available online at:
http://vo.iucaa.ernet.in/$\sim$voi/NIR\_Header.html

\end{abstract}

\section{Introduction}

In the last few years, several population synthesis models have
completely renewed the interest for population analysis. Models by
Vazdekis et al. (1999), Bruzual \& Charlot (2003) and le Borgne et
al. (2004, PEGASE.HR) gained in details with a higher spectral
resolution. The physics of the models has improved substantially,
with the implementation of new evolutionary tracks, in particular
with enhanced Mg/Fe (Thomas \& Maraston 2003). However, the most
remarkable progress concerns the quality of the stellar libraries.

A decade ago, the stellar spectral libraries, even in the visible
region, were lacking in spectral resolution, wavelength coverage
and/or coverage of the parameter space. The best library at that
time was the Jones library (1999) with a resolution R=3000 but with
a restricted wavelength coverage, and a poor knowledge of the
atmospheric parameters of the stars. In the last few years a new
generation of libraries has been published; ELODIE:  with 1500 stars
at R=10000 in the range 390 to 680 nm (Prugniel \& Soubiran 2001;
last version in Prugniel et al. 2007); STELIB: which has a small
number of stars and low resolution but covers all the visible range
(Le Borgne et al. 2003); INDO-US (Valdes et al. 2004) which has both
a large wavelength coverage with a good spectral resolution, R=5000
and MILES: which has a good coverage in wavelength and atmospheric
parameters but still insufficient resolution (Sanchez-Blazquez et
al. 2006).

But the situation is not the same in case of near-infrared. Several
authors have compiled small libraries in K band region (Johnson \&
Me\'ndez 1970 ; Kleinman \& Hall 1986; Lanc\,on \& Rocca-Volmerange
1992; Ali et al. 1995; Hanson et al. 1996; Wallace \& Hinkle 1997).
Most of these libraries are at medium resolution (500-3000). The
most recent is by Ivanov et al. (2004) which contains 218 late type
stars spanning a range of [Fe/H] $\sim$ -2.2 to $\sim$ +0.3 but is
not flux calibrated. In this paper, we present a spectral library of
114 star in K-band at moderate resolution of 22 \AA~ covering larger
range in T$_{eff}$ as compared to Ivanov et al. (2004).

In this paper, Section $\S 2$ describes the observations and related
issues. In section $\S 3$, we describe the basis of selection of the
stars for this library and in section $\S 4$ we describe the data reduction
and calibration procedure. Lastly, in section $\S 5$ we show examples
of some K band spectra and their comparison with the existing
database of Wallace et al. (1997).

\section{OBSERVATIONS}

The database of 114 stars selected in this library were observed in
six different runs from January-April 2003. The details of the log
is shown in Table 1 in which first column gives observing date and
month, column 2 gives the total number of programme stars observed in each run,
last column gives the total number of standard
stars observed in each run. All the observations have been done from
the 1.2 meter Gurushikhar Infrared Telescope (GIRT) of Mt.Abu
Infrared Observatory, India (24$^{0}$39$^{\prime} $
10.9$^{\prime\prime}$N, 72$^{0}$46$^{\prime}$45.9$^{\prime\prime}$E
at an altitude of 1680 meters). The K band long slit spectra were
taken from the NIR Imager/Spectrometer equipped with a
256$\times$256  HgCdTe NICMOS3 array. The slit width corresponds to
2 arc-seconds for the f/13 Cassegrain focus with the slit covering
most of 240 arc-seconds field of view and oriented along North-South
direction in the sky. The reflection grating has 149 lines per mm
and is blazed for H band center wavelength of 1.65 $\mu m$ in the
first order and combined with the slit width of 76 $\mu m$ gives a
moderate resolution of 1000. The exposure time for individual
spectrum ranged from 1 sec to 120 sec depending on the K
magnitude of the program star resulting in S/N ratio of 50 or
better. Two sets of spectra were obtained at two dithered positions
on the array, the typical separation was about 20 arcsec. As the 256
elements of NICMOS3 array in the dispersion axis do not cover the
entire K band, the spectra have been obtained for two grating
settings, denoted as K1 and K2 region. By
combining K1 and K2 region, single K band spectra have been
computed. The details of procedure to acquire the data from the Mt.
Abu observatory is discussed in paper I.

\begin{table}
\caption{Observations Log at GIRT}
\vspace{0.3cm}
\begin{tabular}{ccc} \hline
Dates of Observations & Program Stars &
 Standard Stars\\ \hline

20-24 Jan 03 & 18 & 1 \\
07-12 Feb 03 & 40 & 3 \\
02-04 Mar 03 & 13 & 2 \\
17-20 Mar 03 & 28 & 1 \\
04-07 Apr 03 & 26 & 9 \\
27-30 Apr 03 & 20 & 18 \\ \hline

\end{tabular}
\end{table}

    For a majority of the program stars, we have observed a nearby
main-sequence A type star at nearly same air-mass to minimize the
effects of atmospheric extinction. To optimize the observing
efficiency, a single standard star has been observed whenever some
of the program stars happened to be in the nearby region of the sky.
For the early February and late April 2003 observing runs, late B
type standards have been observed. The list of standard stars that
have been observed are given in Table 2. In this table the standard
star identifier is given in column (1) with right ascension and
declination  for J2000.0 in column (2) and (3) respectively. Columns
(4), (5) and (6) contain the spectro-luminosity class, observed V
magnitude and T$_{eff}$ respectively.

\begin{table}
\caption{Standard Star list with Observational Parameters$^*$}
\vspace{0.3cm}
\begin{tabular}{lcccccl} \hline
HD &   $\alpha$(J2000.0)&
 $\delta$(J2000.0) &  Type &
 V$_{mag}$ &
 {T$_{eff}$} ($^{\circ}$K) \\
(1) & (2) & (3) & (4) &
(5) & (6) \\ \hline

HD71155 &  08 25 39.63 & -03 54 23.13 & A0V   & 3.90 & 9520  \\
HD87901 &  10 08 22.31 & +11 58 01.95 & B7V   & 1.35 & 13000 \\
HD28319 &  04 28 39.74 & +15 52 15.17 & A7III & 3.41 & 8150  \\
HD47105 &  06 37 42.70 & +16 23 57.31 & A0IV  & 1.90 & 9520  \\
HD71155 &  08 25 39.63 & -03 54 23.13 & A0V   & 3.90 & 9520  \\
HD47105 &  06 37 42.70 & +16 23 57.31 & A0IV  & 1.90 & 9520  \\
HD139006&  15 34 41.27 & +26 42 52.90 & A0V   & 2.21 & 9520  \\
HD65456 &  07 57 40.11 & -30 20 04.46 & A2Vvar& 4.79 & 8970  \\
HD97633 &  11 14 14.41 & +15 25 46.45 & A2V   & 3.32 & 8970  \\
HD155125&  17 10 22.69 & -15 43 29.68 & A2.5Va& 2.43 & 8845  \\
HD94601 &  10 55 36.82 & +24 44 59.3  & A1V   & 4.50 & 9230  \\
HD60179 &  07 34 35.9  & +31 53 18    & A1V   & 1.58 & 9230  \\
HD106591&  12 15 25.56 & +57 01 57.42 & A3V   & 3.30 & 8720  \\
HD153808&  17 00 17.37 & +30 55 35.06 & A0V   & 3.91 & 9520  \\
HD103287&  11 53 49.85 & +53 41 41.14 & A0Ve  & 2.43 & 9520  \\
HD130109&  14 46 14.92 & +01 53 34.39 & A0V   & 3.72 & 9520  \\
HD85235 &  09 52 06.36 & +54 03 51.56 & A3IV  & 4.56 & 8720  \\
HD141003&  15 46 11.26 & +15 25 18.57 & A2IV  & 3.66 & 8970  \\
HD79469 &  09 14 21.86 & +02 18 51.41 & B9.5V & 3.88 & 10010 \\
HD118098&  13 34 41.60 & -00 35 44.95 & A3V   & 3.40 & 8720  \\
HD82621 &  09 34 49.43 & +52 03 05.32 & A2V   & 4.48 & 8970  \\
HD141003&  15 46 11.26 & +15 25 18.57 & A2IV  & 3.66 & 8970  \\
HD87737 &  10 07 19.95 & +16 45 45.59 & A0Ib  & 3.51 & 9730  \\
HD141003&  15 46 11.26 & +15 25 18.57 & A2IV  & 3.66 & 8970  \\ \hline
\end{tabular}
\\
$^*$ (2)-(5) From SIMBAD database,(6) From Lang (1992)
\end{table}

      The wavelength calibration has been performed using  OH airglow lines.
Hence, in case of brighter stars when the individual integration
time was less than 120 sec, a separate sky frame was taken with 120
sec exposure time by drifting the star in RA axis by typically about
10 arc-seconds. This enabled the OH airglow lines to register with
reasonably large counts.

\section{SELECTION OF STARS}

While building a spectral library, it is very important  that one
includes various spectral types so that we have a homogeneous and
comprehensive coverage of all possible spectro-luminosity classes.
To optimise the observing efficiency stars upto a magnitude of $\sim$
7 were selected for the present programme. The histogram in Fig. 1 represents the
total number of stars covered in terms of
spectral types (top panel) and Luminosity classes (bottom panel)
The details of number of stars covered in terms of
spectral types per luminosity class is illustrated by the histogram
in Fig. 2. It may be noted that we have covered the HR diagram in
effective temperature and luminosity parameters reasonably well,
although we do not have enough stars for luminosity class II and
main spectral type O. The details of program stars along with the
NIR magnitudes in J,H,K,L \& M are listed in Tables 3, 4 \& 5.
In these Tables,
the first column contains the program star ID, columns (2) to (6)
list the J,H,K,L, \& M magnitudes respectively. The references from
which they have been taken are listed in column 7.

\begin{figure}
\epsfig{file=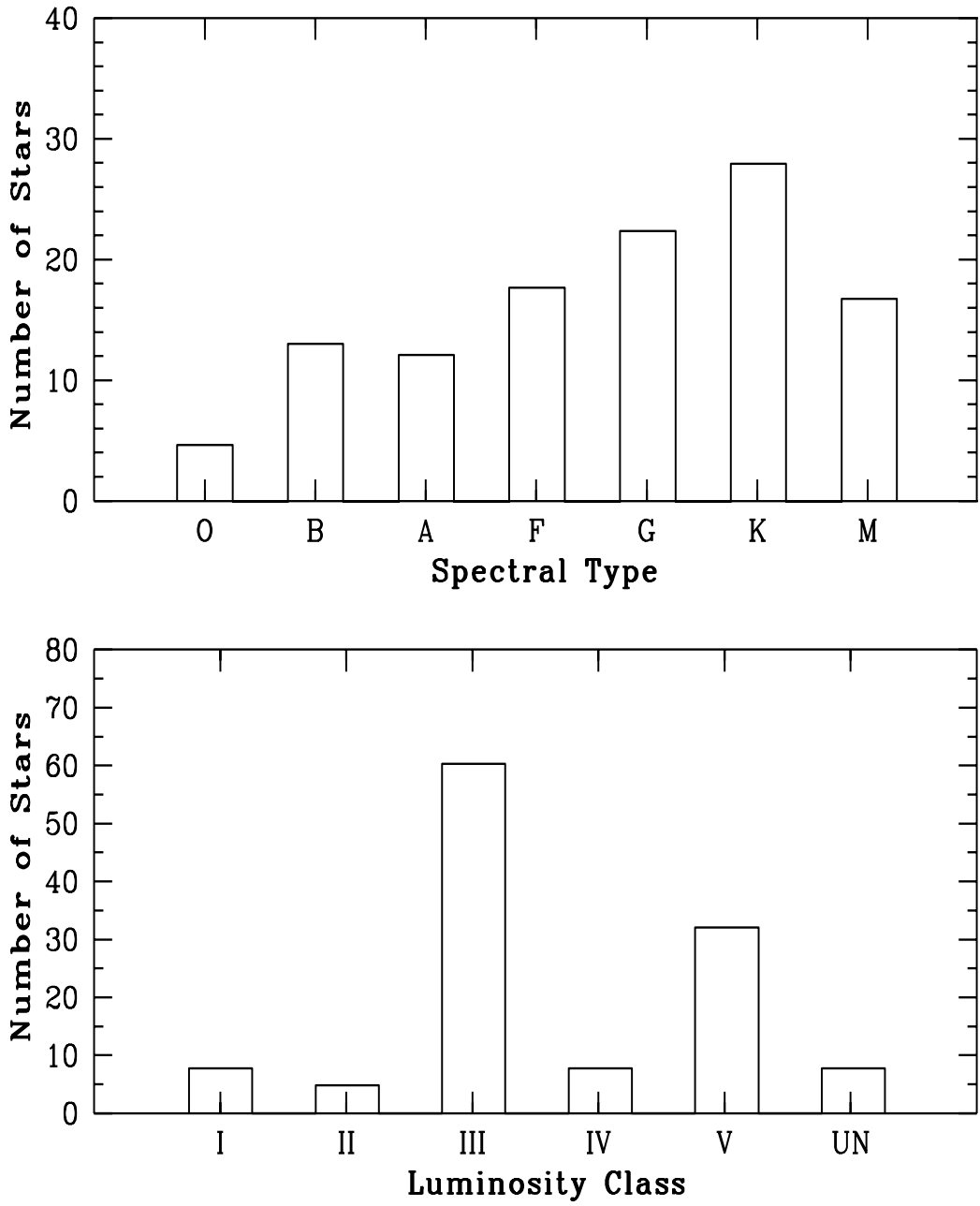,height=18cm,width=15cm}
\caption{Distribution of stars in the database by spectral type and
Luminosity class}
\end{figure}

\begin{figure}
\epsfig{file=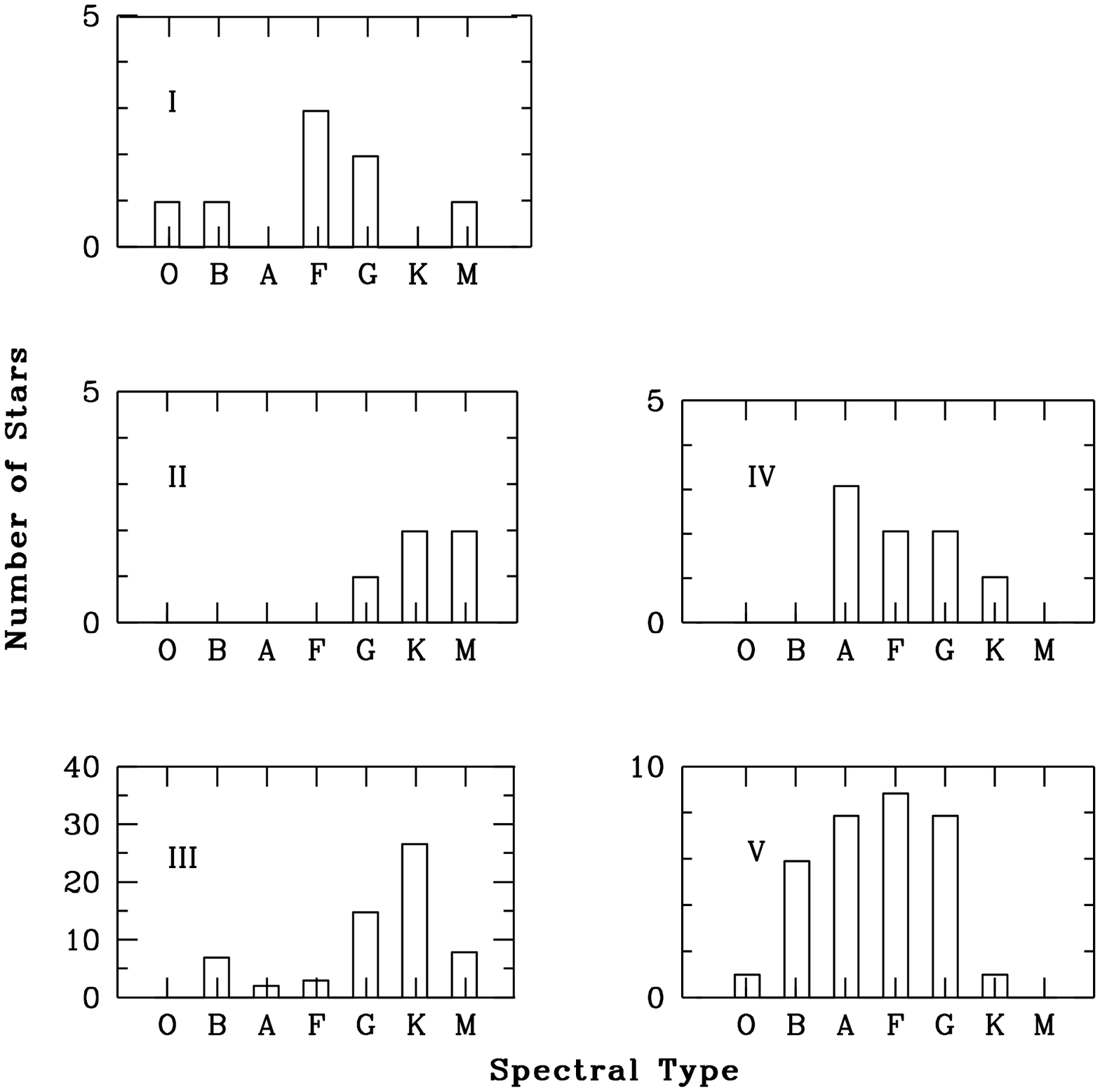,height=18cm,width=15cm}
\caption{Distribution of stars in the database by spectral type per
Luminosity class}
\end{figure}

\begin{table}
\caption{NIR magnitudes of program stars}
\vspace{0.3cm}
\begin{tabular}{ccccccl} \hline
HD & J$_{mag}$ & H$_{mag}$ &
K$_{mag}$ & L$_{mag}$ &
M$_{mag}$ & Reference\\
(1) & (2) & (3) & (4) & (5) & (6) & (7)\\ \hline

HD007927 &      &      &     &       &        & 1997ApJ....111...445 (Wallace) \\
HD008538 &      & 2.30 &     &       &        & 1998ApJ....508...397 (Meyer) \\
HD023475 &      &      &     &       &        & 1997ApJ....111...445 (Wallace) \\
HD025204 & 3.66 & 3.67 &3.66 &  3.65 &  3.71  & 1983A\&AS....51...489 (Koornneef) \\
HD026846 & 2.97 & 2.39 &2.27 &  2.19 &        & 1983A\&AS....51...489 (Koornneef) \\
HD030652 & 2.37 & 2.15 &2.08 &  2.09 &        & 1983A\&AS....51...489 (Koornneef) \\
HD030836 &      & 4.10 &     &       &        & 1998ApJ....508...397 (Meyer) \\
HD035468 & 2.17 & 2.28 &2.32 &  2.34 &  2.36  & 1983A\&AS....51...489 (Koornneef) \\
HD035497 & 1.96 & 2.02 &2.02 &  2.03 &  2.11  & 1983A\&AS....51...489 (Koornneef) \\
HD036673 & 2.05 & 1.92 &1.86 &  1.81 &        & 1983A\&AS....51...489 (Koornneef) \\
HD037128 &      & 2.40 &     &       &        & 1998ApJ....508...397 (Meyer) \\
HD037742 & 2.21 & 2.27 &2.32 &  2.31 &        & 1983A\&AS....51...489 (Koornneef) \\
HD038393 & 2.70 & 2.47 &2.41 &  2.38 &        & 1983A\&AS....51...489 (Koornneef)  \\
HD038858 & 4.82 & 4.50 &4.44 &       &        & 1991A\&AS....91...409 (Bouchet) \\
HD040136 & 3.10 & 2.94 &2.90 &  2.87 &        & 1983A\&AS....51...489 (Koornneef) \\
HD043232 & 1.84 & 1.19 &1.02 &  0.94 &        & 1983A\&AS....51...489 (Koornneef) \\
HD047839 &      & 5.50 &     &       &        & 1998ApJ....508...397 (Meyer) \\
HD048329 &      &      &     &       &        & 1997ApJ....111...445 (Wallace) \\
HD049331 &      &      &     &       &        & 1997ApJ....111...445 (Wallace) \\
HD054605 & 0.77 & 0.51 &0.41 &  0.32 &  0.28  & 1983A\&AS....51...489 (Koornneef) \\
HD054810 & 3.18 & 2.64 &2.53 &  2.47 &        & 1983A\&AS....51...489 (Koornneef) \\
HD056537 &      &      &     &       &        & 1997ApJ....111...445 (Wallace) \\
HD058715 & 1.83 & 1.07 &0.90 &  0.77 &        & 1983A\&AS....51...489 (Koornneef) \\
HD060414 & 1.25 & 0.38 &0.09 &  -0.09&  0.17  & 1983A\&AS....51...489 (Koornneef) \\
HD061935 & 2.28 & 1.77 &1.62 &  1.57 &        & 1983A\&AS....51...489 (Koornneef) \\
HD062576 & 1.74 & 0.96 &0.75 &  0.63 &        & 1983A\&AS....51...489 (Koornneef) \\
HD062721 &      &      &     &       &        & 1997ApJ....111...445 (Wallace) \\
HD063700 & 1.52 & 1.03 &0.89 &  0.81 &        & 1983A\&AS....51...489 (Koornneef) \\
HD065810 & 4.40 & 4.33 &4.32 &  4.31 &        & 1990MNRAS....242...1 (Carter) \\
HD067228 & 4.13 & 3.91 &3.83 &  3.79 &  3.92  & 1983A\&AS....51...489 (Koornneef) \\
HD068312 &      &      &     &       &        & 1997ApJ....111...445 (Wallace) \\
HD070272 &      &      &     &       &        & 1997ApJ....111...445 (Wallace) \\
HD071369 &      &      &     &       &        & 1997ApJ....111...445 (Wallace) \\
HD072094 & 2.45 & 1.64 &1.43 &  1.26 &  1.57  & 1994A\&AS....105...311 (Fluks) \\
HD074918 & 2.80 & 2.33 &2.23 &  2.17 &        & 1983A\&AS....51...489 (Koornneef) \\
HD076943 &      &      &     &       &        & 2004ApJS...152..251 (INDO-US) \\
HD077912 &      &      &     &       &        & 2004ApJS...152..251 (INDO-US) \\
HD085444 & 2.59 & 2.13 &2.02 &  1.97 &        & 1983A\&AS....51...489 (Koornneef) \\
HD085951 & 2.01 & 1.22 &1.01 &  0.85 &  1.18  & 1994A\&AS....105...311 (Fluks) \\
HD086663 & 1.54 & 0.72 &0.50 &  0.34 &  0.66  & 1994A\&AS....105...311 (Fluks) \\
HD088230 &      &      &     &       &        & 2004ApJS...152..251 (INDO-US) \\
HD088284 & 1.99 & 1.51 &1.40 &  1.34 &        & 1983A\&AS....51...489 (Koornneef) \\
HD089010 & 4.78 & 4.47 &4.40 &  4.36 &  4.42  & 1983A\&AS....51...489 (Koornneef) \\
HD089025 &      &2.8   &     &       &        & 1998ApJ....508...397 (Meyer) \\
HD089021 &      & 3.3  &     &       &        & 1998ApJ....508...397 (Meyer) \\
HD089449 &      &      &     &       &        & 2004ApJS...152..251 (INDO-US) \\
HD089490 &      &      &     &       &        & 2004ApJS...152..251 (INDO-US) \\
HD089758 &      &      &     &       &        & 1997ApJ....111...445 (Wallace) \\
\hline
\end{tabular}
\end{table}

\begin{table}
\caption{NIR magnitudes of program stars -- Contd.}
\vspace{0.3cm}
\begin{tabular}{ccccccl} \hline
HD & J$_{mag}$ & H$_{mag}$ &
K$_{mag}$ & L$_{mag}$ &
M$_{mag}$ & Reference\\
(1) & (2) & (3) & (4) & (5) & (6) & (7)\\ \hline

HD090254 & 2.45 & 1.59 &1.36 &  1.20 &  1.48  & 1994A\&AS....105...311 (Fluks) \\
HD090277 &      &      &     &       &        & 2004ApJS...152..251 (INDO-US) \\
HD090432 & 1.31 & 0.56 &0.38 &  0.26 &        & 1983A\&AS....51...489 (Koornneef)\\
HD090610 & 1.81 & 1.07 &0.91 &  0.77 &  1.00  & 1994A\&AS....105...311 (Fluks) \\
HD092125 &      &      &     &       &        & 1997ApJ....111...445 (Wallace) \\
HD092588 &      &      &     &       &        & 2004ApJS...152..251 (INDO-US) \\
HD093813 & 1.07 & 0.42 &0.27 &  0.17 &        & 1983A\&AS....51...489 (Koornneef) \\
HD094264 &      &      &     &       &        & 2004ApJS...152..251 (INDO-US) \\
HD094481 &      &      &     &       &        & 1997ApJ....111...445 (Wallace) \\
HD095418 &      &      &     &       &        & 1997ApJ....111...445 (Wallace) \\
HD097603 & 2.32 & 2.27 &2.27 &  2.29 &        & 1983A\&AS....51...489 (Koornneef) \\
HD097778 &      &      &     &       &        & 1997ApJ....111...445 (Wallace) \\
HD098430 & 1.68 & 1.07 &0.94 &  0.86 &        & 1983A\&AS....51...489 (Koornneef) \\
HD099028 &      &      &     &       &        & 2004ApJS...152..251 (INDO-US) \\
HD099167 &      &      &     &       &        & 2004ApJS...152..251 (INDO-US) \\
HD100407 & 2.01 & 1.58 &1.44 &  1.40 &  1.50  & 1983A\&AS....51...489 (Koornneef) \\
HD100920 &      &      &     &       &        & 2004ApJS...152..251 (INDO-US) \\
HD101501 &      &3.8   &     &       &        & 1998ApJ....508...397 (Meyer) \\
HD102647 &      &2.0   &     &       &        & 1998ApJ....508...397 (Meyer) \\
HD105707 & 0.94 & 0.31 &0.14 &  0.03 &        & 1983A\&AS....51...489 (Koornneef) \\
HD106625 & 2.79 & 2.83 &2.82 &  2.76 &        & 1983A\&AS....51...489 (Koornneef) \\
HD107259 & 3.81 & 3.78 &3.77 &  3.76 &        & 1990MNRAS....242...1 (Carter) \\
HD107328 & 2.95 & 2.32 &2.19 &  2.09 &        & 1983A\&AS....51...489 (Koornneef) \\
HD108767 & 3.06 & 3.08 &3.06 &  3.03 &        & 1983A\&AS....51...489 (Koornneef) \\
HD109358 &      &      &     &       &        & 2004ApJS...152..251 (INDO-US) \\
HD109379 & 1.24 & 0.81 &0.70 &  0.64 &        & 1983A\&AS....51...489 (Koornneef) \\
HD109387 &      &      &     &       &        & 2004ApJS...152..251 (INDO-US) \\
HD110379 & 2.07 & 1.90 &1.86 &  1.84 &        & 1983A\&AS....51...489 (Koornneef) \\
HD111812 & 3.73 & 3.46 &3.36 &  3.29 &  3.34  & 1983A\&AS....51...489 (Koornneef) \\
HD113139 &      &4.1   &     &       &        & 1998ApJ....508...397 (Meyer) \\
HD113226 &      &      &     &       &        & 1997ApJ....111...445 (Wallace) \\
HD113847 &      &      &     &       &        & 2004ApJS...152..251 (INDO-US) \\
HD113996 &      &      &     &       &        & 2004ApJS...152..251 (INDO-US) \\
HD114330 &      &      &     &       &        & 1997ApJ....111...445 (Wallace) \\
HD114961 &      &      &     &       &        & 2004ApJS...152..251 (INDO-US) \\
HD115604 &      &3.9   &     &       &        & 1998ApJ....508...397 (Meyer) \\
HD115892 & 2.73 & 2.74 &2.73 &  2.70 &        & 1983A\&AS....51...489 (Koornneef) \\
HD116656 &      &      &     &--     &        & 1997ApJ....111...445 (Wallace) \\
HD116658 & 1.53 & 1.64 &1.68 &  1.72 & 1.76   & 1983A\&AS....51...489 (Koornneef) \\
HD116870 & 2.62 & 1.81 &1.61 &  1.47 &  1.73  & 1994A\&AS....105...311 (Fluks) \\
HD120315 &      &2.4   &     &       &        & 1998ApJ....508...397 (Meyer) \\
HD120323 & -0.51&-1.39 &-1.66&  -1.84&  -1.4 0& 1983A\&AS....51...489 (Koornneef) \\
HD123139 & 0.42 &-0.09 &-0.21&  -0.31&  -0.2 1& 1983A\&AS....51...489 (Koornneef) \\
HD123299 &      &3.5   &     &       &        & 1998ApJ....508...397 (Meyer) \\
HD123657 &      &      &     &       &        & 2004ApJS...152..251 (INDO-US) \\
HD123934 &      &      &     &       &        & 2004ApJS...152..251 (INDO-US) \\
HD124294 & 1.89 & 1.18 &1.03 &  0.94 &        & 1983A\&AS....51...489 (Koornneef) \\
HD126661 &      &      &     &       &        & 2004ApJS...152..251 (INDO-US) \\ \hline
\end{tabular}
\end{table}

\begin{table}
\caption{NIR magnitudes of program stars -- Contd.}
\vspace{0.3cm}
\begin{tabular}{ccccccl} \hline
HD & J$_{mag}$ & H$_{mag}$ &
K$_{mag}$ & L$_{mag}$ &
M$_{mag}$ & Reference\\
(1) & (2) & (3) & (4) & (5) & (6) & (7)\\ \hline
HD127665 &      &      &     &       &        & 2004ApJS...152..251 (INDO-US) \\
HD129116 & 4.38 & 4.46 &4.52 &  4.56 &        & 1990MNRAS....242...1 (Carter) \\
HD129502 & 3.12 & 2.94 &2.89 &  2.83 &        & 1983A\&AS....51...489 (Koornneef) \\
HD130025 &      &      &     &       &        & 2004ApJS...152..251 (INDO-US) \\
HD130819 & 4.40 & 4.21 &4.16 &  4.13 &  4.13  & 1983A\&AS....51...489 (Koornneef) \\
HD130841 & 2.52 & 2.45 &2.42 &  2.40 &        & 1983A\&AS....51...489 (Koornneef) \\
HD131156 &      &      &     &       &        & 2004ApJS...152..251 (INDO-US) \\
HD134083 &      &      &     &       &        & 2004ApJS...152..251 (INDO-US) \\
HD135722 &      &      &     &       &        & 2004ApJS...152..251 (INDO-US) \\
HD135742 & 2.80 & 2.83 &2.86 &  2.84 &        & 1983A\&AS....51...489 (Koornneef) \\
HD136512 &      &      &     &       &        & 2004ApJS...152..251 (INDO-US) \\
HD141004 & 3.36 & 3.05 &2.99 &       &        & 1991A\&AS....91...409 (Bouchet) \\
HD141714 &      &      &     &       &        & 2004ApJS...152..251 (INDO-US) \\
HD141850 & 2.05 & 1.23 &0.69 &  -0.08&  -0.10 & 1994A\&AS....105...311 (Fluks) \\
HD145328 &      &      &     &       &        & 2004ApJS...152..251 (INDO-US) \\
HD147165 & 2.49 & 2.44 &2.42 &  2.42 &  2.43  & 1983A\&AS....51...489 (Koornneef) \\
HD148513 &      &      &     &       &        & 1990MNRAS....242...1 (Carter) \\
HD161239 &      &      &     &       &        & 1997ApJ....111...445 (Wallace) \\
\hline
\end{tabular}
\end{table}

The detailed criteria for the selection of stars with their
references are discussed in paper I. We have covered a reasonable
region of parameter space in temperature, gravity and metallicity.
Fig. 3 shows the plot of log g vs. T$_{eff}$ for the GIRT stars. The
general trend is quite similar to that of Ivanov et al. (2004)
except for the larger coverage in effective temperature in our case.
Figs. 4 \& 5 shows the plot of [Fe/H] vs. T$_{eff}$ and log g
respectively for the GIRT sample.

\begin{figure}
\epsfig{file=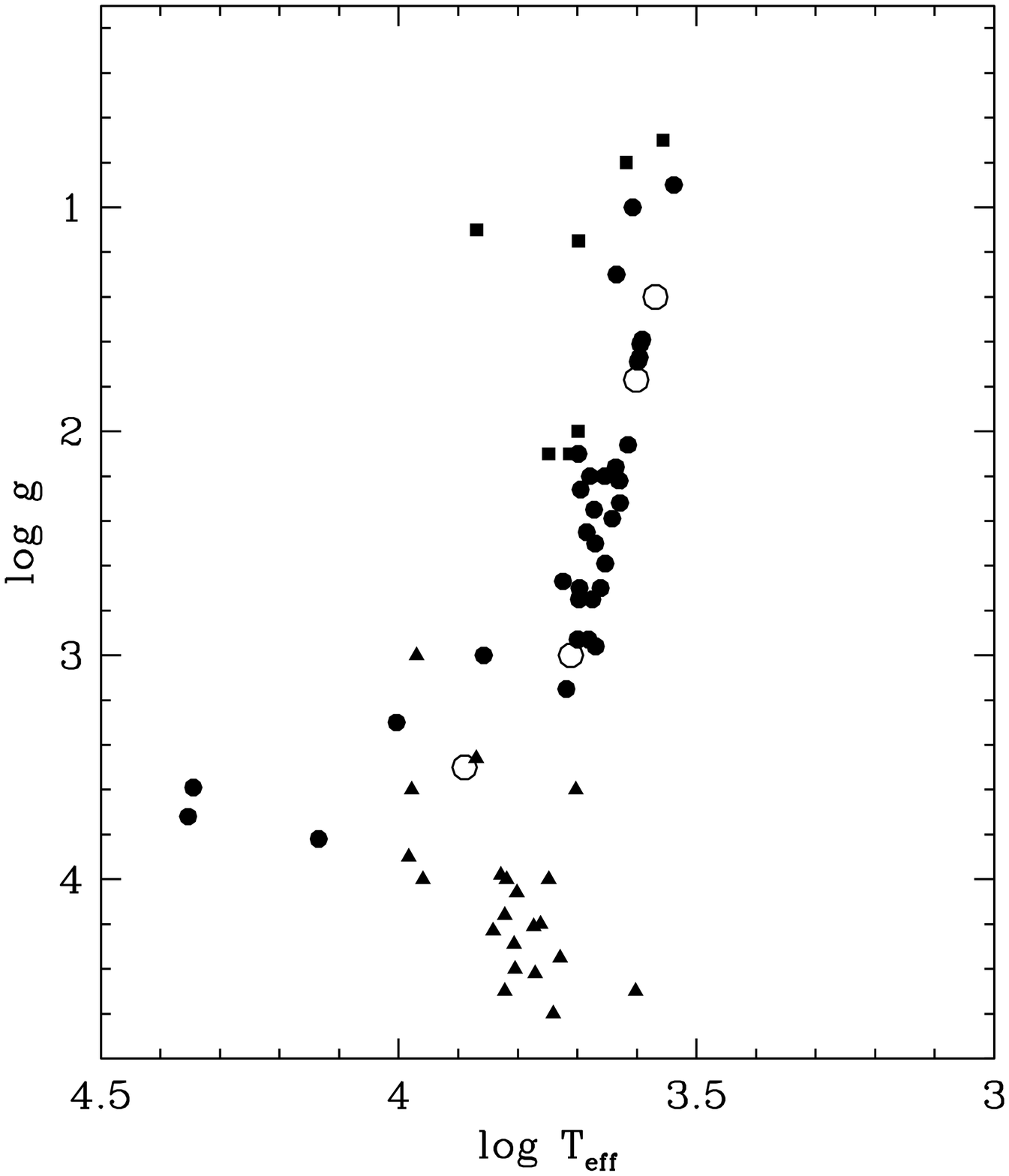,height=18cm,width=15cm}
\caption{ Distribution in the GIRT library: supergiants (I \& II)
with filled squares, giants (III) with filled circles, dwarfs
(IV \& V) with filled triangles and unknown luminosity class with open circles
on surface gravity log g vs. effective temperature T$_{eff}$ plane. }
\end{figure}

\begin{figure}
\epsfig{file=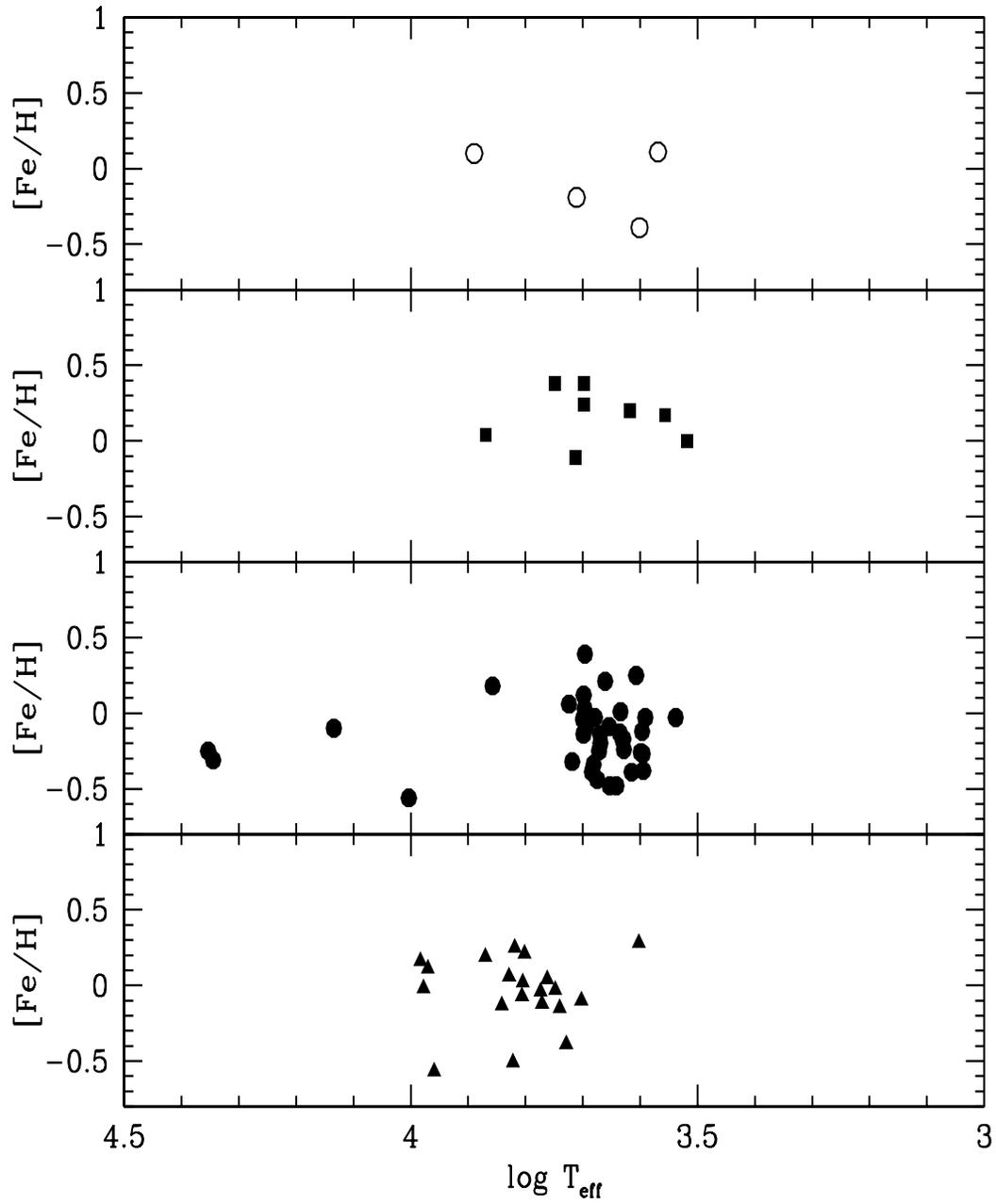,height=18cm,width=15cm}
\caption{Distribution in the GIRT library: unknown luminosity class
with open circles, supergiants (I \& II) with filled squares, giants (III)
with filled circles and dwarfs (IV \& V) with filled triangles ({\it from top to
bottom}) on metallicity [Fe/H] vs. effective temperature T$_{eff}$ plane.}
\end{figure}

\begin{figure}
\epsfig{file=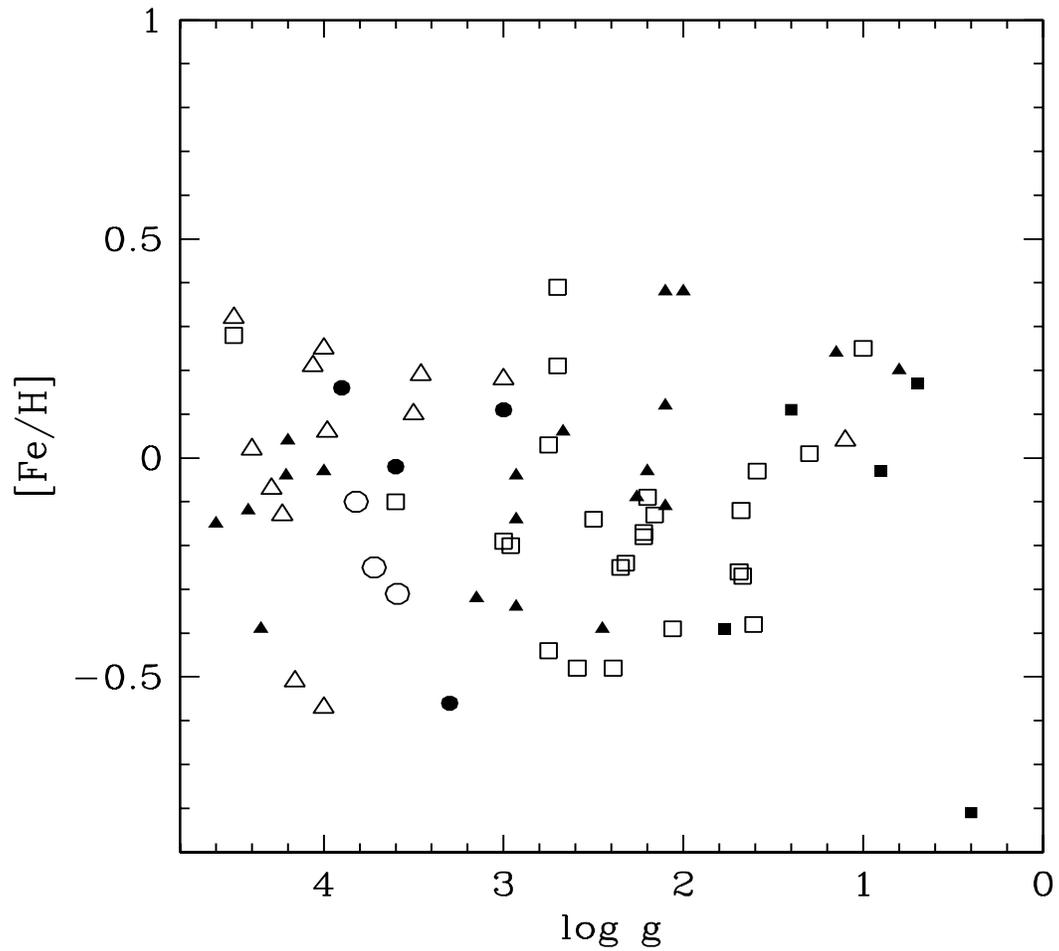,height=18cm,width=15cm}
\caption{Distribution in the GIRT library: Stars of spectral types
B (open circles), A (filled circles) F (open triangles), G (filled triangles),
K (open squares) and M (filled squares) are shown in metallicity
[Fe/H] vs. surface gravity log g plane.}
\end{figure}

\section{DATA REDUCTION AND CALIBRATION}

 The near infrared spectral data reduction is similar to that of optical
data reduction with minor differences. The presence of strong
telluric emission lines and varying atmospheric transmission due to
changing water vapour content necessitates observation of standard
star spectra at similar airmass soon after the program star
observation. The whole process of near infrared long slit spectra
reduction can be separated into a few major steps, viz., (i)
pre-processing (ii) spectrum extraction (iii) wavelength calibration
(iv) atmospheric transmission and instrument response determination
using standard star data (v) continuum fitting and (vi) radial
velocity correction. We have used standard tasks available in
software package IRAF \footnote {IRAF is distributed by National
Optical Astronomy Observatories, which are operated by the
Association of Universities for Research in Astronomy, Inc., under
cooperative agreement with the National Science Foundation.} for
data reduction. As discussed in \S2 we have two sets of frames at
two different locations of the detector. The availability of these
two sets of spectra is utilized to remove the dark counts and the
large sky background at near infrared wavelengths. This is
accomplished by taking the difference of  spectra obtained at two
different locations on the detector. As there is no autoguider on
the telescope, the frames with maximum counts in two positions are
selected for data reduction. We thus have two difference frames for
extraction of the spectrum. The detail of the each task and its
significance in the data reduction is discussed in paper I. As
discussed in \S2 registering atmospheric OH airglow lines in the
frame is important which helps for wavelength calibration of the
stellar spectra. The list of registered OH airglow lines are given
in Table 6. The IRAF task {\it{identify}} is used for this purpose.
The IRAF task {\it{refspec}} is used to specify the appropriate
wavelength calibrated spectrum for the stellar spectra extracted
through {\it{apall}} task. The IRAF task {\it{dispcor}} was used to
set the wavelength calibration for the stellar spectra.

\begin{table}
\caption{Line List of atmospheric OH airglow emissions used
for wavelength calibrations in K-band}
\vspace{0.3cm}
\begin{tabular}{cc} \hline
Sr. Number & Wavelength in \AA \\ \hline
1  & 20008.2  \\
2  & 20275.9  \\
3  & 20412.7  \\
4  & 20563.6  \\
5  & 20729.0  \\
6  & 20909.6  \\
7  & 21507.3  \\
8  & 21802.2  \\
9  & 21956.0  \\
10  & 22125.5  \\
11  & 22312.7  \\
12  & 22516.7  \\ \hline
\end{tabular}
\end{table}

The effects like atmospheric transmission effects and the instrument effects
(filter transmission and wavelength dependence of detector quantum efficiency)
can be be removed by taking the ratio of the program star spectrum
with that of a standard star spectrum observed under similar conditions and
subsequently multiplying with a model synthetic spectrum for the standard star.
We have selected bright A and late B type with T$_{eff} \approx $ 10000 K because at this
temperature only neutral hydrogen lines will be present and no metallic lines will
survive in the NIR spectral region. Table 2 lists standard stars used for
the purpose of taking ratios. The stellar absorption feature due to hydrogen namely
the Brackett gamma line was removed before taking the ratio. The program star flux is divided
by the corresponding standard star flux and in this process the modulation due to telluric
features, atmospheric extinction and instrumental effects cancels out. The resultant function from
this division is multiplied with a corresponding blackbody flux distribution
at the temperature corresponding to the standard star.
It may be noted that unlike many of the spectral libraries published earlier
the spectra presented here have been continuum shape corrected
to their respective effective temperatures.

    As mentioned in Section 2, majority of the spectra were obtained at two
positions (K1 and K2) of the grating to cover the entire K band region.
While doing the data reduction it was observed that registration of OH airglow lines at
higher wavelength (K2 region) is quite poor and difficult to use for wavelength calibration.
The slightly different placement of the grating in K2 setting resulted in non-uniform
coverage of longer wavelengths for each star. Hence most of the data in this
library spans the earlier part of the K band with slightly different wavelength coverage
corresponding to the K1 setting. The detail of the wavelength coverage for
each star is given in Tables 7, 8 \& 9.
In Tables 7, 8 \& 9, the first and second column
contain the star ID, Columns (3) and (4) contain the right ascension
(J2000.0) and declination (J2000.0) respectively Column (5) gives the apparent $V$ magnitude.
The column (6) gives the corresponding standard star ID used for data reduction while last column
gives the available GIRT spectral region for the same star.

\begin{table}
\caption{Observational Parameters (from SIMBAD) of program stars}
\vspace{0.3cm}
\begin{tabular}{ccccccc} \hline
HD & HR & $\alpha$(J2000.0) &
$\delta$(J2000.0) &
V$_{mag}$ & Standard Star & NIR Coverage\\
& & & & & & $\mu m)$\\
(1) & (2) & (3) & (4) & (5) & (6) & (7) \\ \hline
HD007927 & HR382  & 01 20 04.91 & +58 13 53.79  &  5.01 & HR3314 & 2.04-2.26 \\
HD008538 & HR403  & 01 25 48.95 & +60 14 07.01  &  2.68 & HR3314 & 2.03-2.21 \\
HD023475 & HR1155 & 03 49 31.28 & +65 31 33.50  &  4.47 & HR2421 & 2.05-2.17 \\
HD025204 & HR1239 & 04 00 40.81 & +12 29 25.24  &  3.40 & HR2421 & 2.05-2.19 \\
HD026846 & HR1318 & 04 14 23.68 & -10 15 22.61  &  4.90 & HR2421 & 2.05-2.17 \\
HD030652 & HR1543 & 04 49 50.41 & +06 57 40.59  &  3.19 & HR3314 & 2.03-2.20 \\
HD030836 & HR1552 & 04 51 12.36 & +05 36 18.37  &  4.47 & HR2421 & 2.05-2.19 \\
HD035468 & HR1790 & 05 25 07.86 & +06 20 58.92  &  1.62 & HR3314 & 2.05-2.22 \\
HD03549- & HR1791 & 05 26 17.51 & +28 36 26.82  &  1.68 & HR3314 & 2.03-2.21 \\
HD036673 & HR1865 & 05 32 43.81 & -17 49 20.23  &  2.59 & HR2421 & 2.04-2.18 \\
HD037128 & HR1903 & 05 36 12.81 & -01 12 06.91  &  1.70 & HR3314 & 2.09-2.20 \\
HD037742 & HR1948 & 05 40 45.53 & -01 56 33.50  &  1.70 & HR3314 & 2.08-2.24 \\
HD038393 & HR1983 & 05 44 27.79 & -22 26 54.17  &  3.60 & HR2421 & 2.08-2.23 \\
HD038858 & HR2007 & 05 48 34.94 & -04 05 40.73  &  5.97 & HR2421 & 2.03-2.15 \\
HD040136 & HR2085 & 05 56 24.29 & -14 10 03.72  &  3.71 & HR1412 & 2.05-2.19 \\
HD043232 & HR2227 & 06 14 51.33 & -06 14 29.19  &  3.98 & HR3314 & 2.08-2.19 \\
HD047839 & HR2456 & 06 40 58.66 & +09 53 44.71  &  4.66 & HR3314 & 2.08-2.25 \\
HD048329 & HR2473 & 06 43 55.92 & +25 07 52.04  &  3.01 & HR2421 & 2.05-2.19 \\
HD049331 & HR2508 & 06 47 37.22 & -08 59 54.60  &  5.10 & HR3982 & 2.05-2.19 \\
HD054605 & HR2693 & 07 08 23.48 & -26 23 35.51  &  1.84 & HR5793 & 2.04-2.19 \\
HD054810 & HR2701 & 07 10 13.68 & -04 14 13.58  &  4.92 & HR3314 & 2.06-2.20 \\
HD056537 & HR2763 & 07 18 05.57 & +16 32 25.37  &  3.58 & HR2421 & 2.04-2.18 \\
HD058715 & HR2845 & 07 27 09.04 & +08 17 21.53  &  2.88 & HR2421 & 2.04-2.23 \\
HD060414 & HR2902 & 07 33 47.96 & -14 31 26.01  &  4.97 & HR1412 & 2.05-2.18 \\
HD061935 & HR2970 & 07 41 14.83 & -09 33 04.07  &  3.93 & HR2421 & 2.05-2.17 \\
HD062576 & HR2993 & 07 43 32.38 & -28 24 39.18  &  4.62 & HR2891 & 2.03-2.19 \\
HD062721 & HR3003 & 07 46 07.44 & +18 30 36.15  &  4.88 & HR2421 & 2.03-2.19 \\
HD063700 & HR3045 & 07 49 17.65 & -24 51 35.22  &  3.33 & HR3113 & 2.05-2.19 \\
HD065810 & HR3131 & 07 59 52.05 & -18 23 57.22  &  4.61 & HR5793 & 2.04-2.20 \\
HD067228 & HR3176 & 08 07 45.85 & +21 34 54.53  &  5.30 & HR2421 & 2.05-2.18 \\
HD068312 & HR3212 & 08 11 33.00 & -07 46 21.14  &  5.35 & HR5793 & 2.03-2.20 \\
HD070272 & HR3275 & 08 22 50.10 & +43 11 17.27  &  4.25 & HR3982 & 2.05-2.17 \\
HD071369 & HR3323 & 08 30 15.87 & +60 43 05.40  &  3.37 & HR2421 & 2.05-2.18 \\
HD072094 & HR3357 & 08 31 35.70 & +18 05 40.00  &  5.33 & HR5793 & 2.05-2.19 \\
HD074918 & HR3484 & 08 46 22.53 & -13 32 51.79  &  4.32 & HR3314 & 2.04-2.21 \\
HD076943 & HR3579 & 09 00 38.40 & +41 46 58.00  &  3.90 & HR2891 & 2.03-2.18 \\
HD077912 & HR3612 & 09 06 31.80 & +38 27 08.00  &  4.50 & HR3894 & 2.05-2.21 \\
HD085444 & HR3903 & 09 51 28.69 & -14 50 47.77  &  4.11 & HR2421 & 2.05-2.18 \\
HD085951 & HR3923 & 09 54 52.20 & -19 00 34.00  &  4.93 & HR5793 & 2.05-2.20 \\
HD086663 & HR3950 & 10 00 12.80 & +08 02 39.00  &  4.64 & HR3799 & 2.03-2.22 \\
HD088230 &        & 10 11 22.14 & +49 27 15.25  &  6.61 & HR3665 & 2.05-2.20 \\
HD088284 & HR3994 & 10 10 35.27 & -12 21 14.69  &  3.61 & HR5793 & 2.05-2.21 \\
HD089010 & HR4030 & 10 16 32.28 & +23 30 11.14  &  5.90 & HR5793 & 2.05-2.18 \\
HD089025 & HR4031 & 10 16 41.41 & +23 25 02.31  &  3.44 & HR2421 & 2.03-2.19 \\
HD089021 & HR4033 & 10 17 05.79 & +42 54 51.71  &  3.44 & HR2421 & 2.05-2.18 \\
HD089449 & HR4054 & 10 19 44.10 & +19 28 15.00  &  4.70 & HR4259 & 2.05-2.20 \\
HD089490 & HR4059 & 10 19 32.20 & -05 06 21.00  &  6.30 & HR4359 & 2.05-2.18 \\
HD089758 & HR4069 & 10 22 19.74 & +41 29 58.25  &  3.06 & HR2421 & 2.03-2.18 \\
\hline
\end{tabular}
\end{table}

\begin{table}
\caption{Observational Parameters (from SIMBAD) of program stars -- Contd.}
\vspace{0.3cm}
\begin{tabular}{ccccccc} \hline
HD & HR & $\alpha$(J2000.0) &
$\delta$(J2000.0) &
V$_{mag}$ & Standard Star & NIR coverage \\
& & & & & & $(\mu$ m)\\
(1) & (2) & (3) & (4) & (5) & (6) & (7)\\ \hline

HD090254 & HR4088 & 10 25 15.20 & +08 47 25.00  &  5.59 & HR3799 & 2.03-2.20 \\
HD090277 & HR4090 & 10 25 59.90 & +33 47 46.00  &  4.70 & HR4554 & 2.03-2.19 \\
HD090432 & HR4094 & 10 26 05.42 & -16 50 10.64  &  3.83 & HR1412 & 2.03-2.20 \\
HD090610 & HR4104 & 10 27 09.10 & -31 04 04.00  &  4.27 & HR4660 & 2.05-2.21 \\
HD092125 & HR4166 & 10 38 43.21 & +31 58 34.45  &  4.68 & HR5793 & 2.03-2.21 \\
HD092588 & HR4182 & 10 41 24.62 & -01 44 23.50  &  6.26 & HR4359 & 2.04-2.20 \\
HD093813 & HR4232 & 10 49 37.48 & -16 11 37.13  &  3.11 & HR5793 & 2.04-2.20 \\
HD094264 & HR4247 & 10 53 18.33 & +34 13 07.30  &  3.03 & HR4554 & 2.04-2.18 \\
HD094481 & HR4255 & 10 54 17.77 & -13 45 28.92  &  5.66 & HR5793 & 2.03-2.17 \\
HD095418 & HR4295 & 11 01 50.47 & +56 22 56.73  &  2.34 & HR5793 & 2.04-2.20 \\
HD097603 & HR4357 & 11 14 06.50 & +20 31 25.38  &  2.56 & HR5793 & 2.03-2.18 \\
HD097778 & HR4362 & 11 15 12.22 & +23 05 43.80  &  4.58 & HR2421 & 2.04-2.20 \\
HD098430 & HR4382 & 11 19 20.44 & -14 46 42.74  &  3.50 & HR5793 & 2.04-2.19 \\
HD099028 & HR4399 & 11 23 55.50 & +10 31 45.00  &  3.90 & HR4259 & 2.05-2.19 \\
HD099167 & HR4402 & 11 24 36.62 & -10 51 34.90  &  4.83 & HR4357 & 2.04-2.19 \\
HD100407 & HR4450 & 11 33 00.11 & -31 51 27.45  &  3.54 & HR5793 & 2.05-2.20 \\
HD100920 & HR4471 & 11 36 57.02 & -00 49 26.00  &  4.30 & HR4554 & 2.04-2.18 \\
HD101501 & HR4496 & 11 41 03.01 & +34 12 05.88  &  5.32 & HR2421 & 2.04-2.18 \\
HD102647 & HR4534 & 11 49 03.57 & +14 34 19.41  &  2.14 & HR2421 & 2.04-2.18 \\
HD105707 & HR4630 & 12 10 07.48 & -22 37 11.15  &  3.01 & HR5793 & 2.05-2.18 \\
HD106625 & HR4662 & 12 15 48.37 & -17 32 30.94  &  2.59 & HR2421 & 2.05-2.19 \\
HD107259 & HR4689 & 12 19 54.35 & -00 40 00.49  &  3.89 & HR5793 & 2.05-2.19 \\
HD107328 & HR4695 & 12 20 20.98 & +03 18 45.26  &  2.06 & HR2421 & 2.07-2.19 \\
HD108767 & HR4757 & 12 29 51.85 & -16 30 55.55  &  2.95 & HR2421 & 2.05-2.19 \\
HD109358 & HR4785 & 12 33 47.64 & +41 21 12.00  &  4.26 & HR4660 & 2.05-2.19 \\
HD109379 & HR4786 & 12 34 23.23 & -23 23 48.33  &  2.65 & HR5793 & 2.04-2.19 \\
HD109387 & HR4787 & 12 33 29.00 & +69 47 18.00  &  3.80 & HR4554 & 2.05-2.18 \\
HD110379 & HR4825 & 12 41 39.60 & -01 26 57.90  &  3.65 & HR3314 & 2.04-2.19 \\
HD111812 & HR4883 & 12 51 41.92 & +27 32 26.56  &  4.93 & HR3982 & 2.04-2.18 \\
HD113139 & HR4931 & 13 00 43.59 & +56 21 58.81  &  4.93 & HR2421 & 2.03-2.18 \\
HD113226 & HR4932 & 13 02 10.59 & +10 57 32.94  &  2.83 & HR3982 & 2.03-2.17 \\
HD113847 & HR4945 & 13 05 52.30 & +45 16 07.00  &  5.60 & HR4660 & 2.05-2.20 \\
HD113996 & HR4954 & 13 07 10.70 & +27 37 29.00  &  4.80 & HR5867 & 2.05-2.19 \\
HD114330 & HR4963 & 13 09 56.99 & -05 32 20.43  &  4.38 & HR5793 & 2.05-2.20 \\
HD114961 &        & 13 14 04.45 & -02 48 24.70  &  7.02 & HR5867 & 2.05-2.19 \\
HD115604 & HR5017 & 13 17 32.54 & +40 34 21.38  &  4.72 & HR3314 & 2.04-2.19 \\
HD115892 & HR5028 & 13 20 35.81 & -36 42 44.26  &  2.70 & HR2421 & 2.05-2.18 \\
HD116656 & HR5054 & 13 23 55.54 & +54 55 31.30  &  2.70 & HR2421 & 2.05-2.20 \\
HD116658 & HR5056 & 13 25 11.57 & -11 09 40.75  &  1.04 & HR2421 & 2.05-2.18 \\
HD116870 & HR5064 & 13 26 43.16 & -12 42 27.59  &  5.27 & HR5793 & 2.05-2.22 \\
HD120315 & HR5191 & 13 47 32.43 & +49 18 47.75  &  1.86 & HR2421 & 2.05-2.19 \\
HD120323 & HR5192 & 13 49 26.72 & -34 27 02.79  &  4.19 & HR5793 & 2.05-2.18 \\
HD123139 & HR5288 & 14 06 40.94 & -36 22 11.83  &  2.06 & HR5893 & 2.05-2.18 \\
HD123299 & HR5291 & 14 04 23.34 & +64 22 33.06  &  3.65 & HR2421 & 2.05-2.18 \\
HD123657 & HR5299 & 14 07 55.65 & +43 51 17.30  &  5.27 & HR5511 & 2.03-2.18 \\
HD123934 & HR5301 & 14 10 50.50 & -16 18 07.00  &  4.90 & HR4259 & 2.05-2.18 \\
HD124294 & HR5315 & 14 12 53.74 & -10 16 25.32  &  4.19 & HR2421 & 2.05-2.19 \\
\hline
\end{tabular}
\end{table}

\begin{table}
\caption{Observational Parameters (from SIMBAD) of program stars -- Contd.}
\vspace{0.3cm}
\begin{tabular}{ccccccc} \hline
HD & HR & $\alpha$(J2000.0) &
$\delta$(J2000.0) &
V$_{mag}$ & Standard Star & NIR coverage \\
& & & & & & $(\mu$ m)\\
(1) & (2) & (3) & (4) & (5) & (6) & (7)\\ \hline
HD126661 & HR5405 & 14 26 27.36 & +19 13 36.83  &  5.39 & HR3314 & 2.05-2.19 \\
HD127665 & HR5429 & 14 31 50.13 & +30 22 11.00  &  3.58 & HR6324 & 2.05-2.19 \\
HD129116 & HR5471 & 14 41 57.59 & -37 47 36.59  &  3.98 & HR5793 & 2.05-2.20 \\
HD129502 & HR5487 & 14 43 03.62 & -05 39 29.54  &  3.90 & HR5793 & 2.05-2.18 \\
HD130025 & HR5507 & 14 45 20.70 & +18 53 05.00  &  6.59 & HR5107 & 2.06-2.20 \\
HD130819 & HR5530 & 14 50 41.18 & -15 59 50.05  &  5.15 & HR5793 & 2.05-2.20 \\
HD130841 & HR5531 & 14 50 52.71 & -16 02 30.40  &  2.75 & HR5793 & 2.05-2.20 \\
HD131156 & HR5544 & 14 51 23.30 & +19 06 04.00  &  4.50 & HR6324 & 2.05-2.18 \\
HD134083 & HR5634 & 15 07 17.34 & +24 52 17.00  &  4.93 & HR5867 & 2.05-2.19 \\
HD135722 & HR5681 & 15 15 29.77 & +33 18 58.70  &  3.47 & HR6324 & 2.05-2.19 \\
HD135742 & HR5685 & 15 17 00.41 & -09 22 58.50  &  2.60 & HR3314 & 2.05-2.19 \\
HD136512 & HR5709 & 15 20 08.94 & +29 37 00.00  &  5.51 & HR5511 & 2.05-2.18 \\
HD141004 & HR5868 & 15 46 26.61 & +07 21 11.06  &  4.43 & HR6378 & 2.05-2.19 \\
HD141714 & HR5889 & 15 49 35.88 & +26 04 09.00  &  4.63 & HR5511 & 2.05-2.19 \\
HD141850 & HR5894 & 15 50 41.70 & +15 08 01.00  &  7.10 & HR6324 & 2.05-2.19 \\
HD145328 & HR6018 & 16 08 58.45 & +36 29 10.30  &  4.76 & HR5107 & 2.05-2.19 \\
HD147165 & HR6084 & 16 21 11.31 & -25 35 34.06  &  2.91 & HR6324 & 2.05-2.19 \\
HD148513 & HR6136 & 16 28 33.98 & +00 39 54.00  &  5.90 & HR6378 & 2.05-2.19 \\
HD161239 & HR6608 & 17 43 21.56 & +24 19 40.15  &  5.74 & HR5867 & 2.04-2.20 \\
\hline
\end{tabular}
\end{table}

\section{SPECTRAL LIBRARY}

The NIR K band spectral library of 114 stars is available in the format
of reduced ASCII tables with wavelength versus flux at a spectral resolution
of 1000 at 5 \AA~ binning. The main goal of this paper is to make this library
available for variety of investigators working in the NIR region.
Thus, the complete library can be downloaded from the website:

http://vo.iucaa.ernet.in/$\sim$voi/NIR\_Header.html

The essential information of each star in the database is summarized in
Tables 3, 4 \& 5 and 7, 8 \& 9 as observational parameters and in Tables
10, 11 \& 12 as
physical parameters. The contents of Tables 3, 4 \& 5 has been mentioned in
section \S3 and content of Tables 7, 8 \& 9 has been mentioned in
section \S4.
Tables 10, 11 \& 12 contains the star ID in the first column. Columns (2),
(3) and (4) give the main spectral type, luminosity class and effective temperature
respectively. Columns (5) and (6) give the log g and [Fe/H] values
respectively. The last column gives the references from which the
physical parameters have been obtained.

\begin{table}
\caption{Physical Parameters of program stars}
\vspace{0.3cm}
\begin{tabular}{ccccccl} \hline
HD & Spectral & Luminosity &
T$_{eff}$($^{\circ}$K)& log$_{10}$(g) & (Fe/H) &Reference\\
 & Type & Class &
& & & \\
(1) & (2)& (3) & (4) & (5) & (6) & (7) \\ \hline
HD007927 & F0   & Ia    &       &      &       &         \\
HD008538 & A5   & III   & 8090  &      &       & 1995A\&AS...110..553 (Sokolov)  \\
HD023475 & M2   & IIab  &       &      &       &         \\
HD025204 & B3   & V     &       &      &       &         \\
HD026846 & K3   & III   & 4582  & 2.70 & 0.21  & 1997A\&AS...124..299C (Cayrel) \\
HD030652 & F6   & V     & 6380  & 4.40 & 0.02  & 2004ApJS...152..251 (INDO-US) \\
HD030836 & B2   & III   & 22120 & 3.59 & -0.31 & 1997A\&AS...124..299C (Cayrel) \\
HD035468 & B2   & III   & 22570 & 3.72 & -0.25 & 2004ApJS...152..251 (INDO-US) \\
HD035497 & B7   & III   & 13622 & 3.80 & -0.10 & 2004ApJS...152..251 (INDO-US) \\
HD036673 & F0   & Ib    & 7400  & 1.10 & 0.04  & 2004ApJS...152..251 (INDO-US) \\
HD037128 & B0   & Iab   &       &      &       &         \\
HD037742 & O9   & Iab   &       &      &       &         \\
HD038393 & F7   & V     & 6398  & 4.29 & -0.07 & 1997A\&AS...124..299C (Cayrel) \\
HD038858 & G4   & V     &       &      &       &         \\
HD040136 & F1   & V     & 6939  & 4.23 & -0.13 & 2004ApJS...152..251 (INDO-US) \\
HD043232 & K1.5 & III   & 4270  & 2.22 & -0.18 & 2004ApJS...152..251 (INDO-US) \\
HD047839 & O7   & Ve    &       &      &       &         \\
HD048329 & G8   & Ib    & 4150  & 0.80 & 0.20  & 2004ApJS...152..251 (INDO-US) \\
HD049331 & M1   & Iab   & 3600  & 0.70 & 0.17  & 1997A\&AS...124..299C (Cayrel) \\
HD054605 & F8   & Iab   &       &      &       &     \\
HD054810 & K0   & III   & 4697  & 2.35 & -0.25 & 2004ApJS...151..387 (Ivanov) \\
HD056537 & A3   & V     &       &      &       &     \\
HD058715 & B8   & Ve    & 11710 &      &       & 1995A\&AS...110..553 (Sokolov)  \\
HD060414 & M2   & III   &       &      &       &         \\
HD061935 & G9   & III   & 4776  & 2.20 & -0.03 & 2004ApJS...151..387 (Ivanov) \\
HD062576 & K3   & III   & 4308  & 1.30 & 0.01  & 1997A\&AS...124..299C (Cayrel) \\
HD062721 & K4   & III   & 3940  & 1.67 & -0.27 & 1997A\&AS...124..299C (Cayrel) \\
HD063700 & G6   & Ia    & 4990  & 1.15 & 0.24  & 1997A\&AS...124..299C (Cayrel) \\
HD065810 & A1   & V     &       &      &       &         \\
HD067228 & G1   & IV    & 5779  & 4.20 & 0.04  & 2004ApJS...152..251 (INDO-US) \\
HD068312 & G6   & III   &       &      &       &     \\
HD070272 & K4.5 & III   & 3900  & 1.59 & -0.03 & 1997A\&AS...124..299C (Cayrel) \\
HD071369 & G5   & III   & 5300  & 2.67 & 0.06  & 2004ApJS...152..251 (INDO-US) \\
HD072094 & K5   & III   &       &      &       &     \\
HD074918 & G8   & III   & 4950  & 2.26 & -0.09 & 1997A\&AS...124..299C (Cayrel) \\
HD076943 & F3   & V     & 6590  & 4.00 & 0.25  & 2004ApJS...152..251 (INDO-US) \\
HD077912 & G7   & Ib-II & 5000  & 2.00 & 0.38  & 2004ApJS...152..251 (INDO-US) \\
HD085444 & G6   & III   & 5000  & 2.93 & -0.14 & 2004ApJS...152..251 (INDO-US) \\
HD085951 & k5   & III   &       &      &       &        \\
HD086663 & M2   & III   &       &      &       &        \\
HD088230 & K8   & V     & 4000  & 4.50 & 0.28  & 1997A\&AS...124..299C (Cayrel)\\
HD088284 & K0   & III   & 4971  & 2.70 & 0.39  & 2004ApJS...151..387 (Ivanov)\\
HD089010 & G1.5 & IV    & 5600  & 4.00 & -0.03 & 2004ApJS...152..251 (INDO-US)\\
HD089025 & F0   & III   &       &      &       &    \\
HD089021 & A1   & IV    &       &      &       &        \\
HD089449 & F6   & IV    & 6333  & 4.06 & 0.21  & 2004ApJS...152..251 (INDO-US)\\
HD089490 & K0   &       &       &      &       &     \\
HD089758 & M0   & III   &       &      &       &      \\
\hline
\end{tabular}
\end{table}

\begin{table}
\caption{Physical Parameters of program stars -- Contd.}
\vspace{0.3cm}
\begin{tabular}{ccccccl} \hline
HD & Spectral & Luminosity &
T$_{eff}$($^{\circ}$K)& log$_{10}$(g) & (Fe/H) &Reference\\
 & Type & Class &
& & & \\
(1) & (2)& (3) & (4) & (5) & (6) & (7) \\ \hline

HD090254 & M3   & III   & 3706  & 1.40 & 0.11  & 1997A\&AS...124..299C (Cayrel) \\
HD090277 & F0   & V     & 7412  & 3.46 & 0.19  & 2004ApJS...152..251 (INDO-US)\\
HD090432 & K4   & III   & 3950  & 1.68 & -0.12 & 1997A\&AS...124..299C (Cayrel)\\
HD090610 & K4   & III   & 3990  & 1.77 & -0.39 & 1997A\&AS...124..299C (Cayrel) \\
HD092125 & G2.5 & IIa   & 5600  & 2.10 & 0.38  & 2004ApJS...152..251 (INDO-US)\\
HD092588 & K1   & IV    & 5044  & 3.60 & -0.10 & 2004ApJS...152..251 (INDO-US)\\
HD093813 & K0   & III   & 4250  & 2.32 & -0.24 & 2004ApJS...152..251 (INDO-US) \\
HD094264 & K0   & III   & 4670  & 2.96 & -0.20 & 2004ApJS...152..251 (INDO-US)\\
HD094481 & K0   & II+.. &       &      &       &    \\
HD095418 & A1   & V     & 9620  & 3.90 & 0.16  & 2004ApJS...152..251 (INDO-US)\\
HD097603 & A4   & V     & 8080  &      &       & 1995A\&AS...110..553 (Sokolov) \\
HD097778 & M3   & IIb   & 3300  &      & 0.00  & 2004ApJS...151..387 (Ivanov)\\
HD098430 & K0   & III   & 4500  & 2.59 & -0.48 & 2004ApJS...152..251 (INDO-US)\\
HD099028 & F1   & IV    & 6739  & 3.98 & 0.06  & 2004ApJS...152..251 (INDO-US)\\
HD099167 & K5   & III   & 3930  & 1.61 & -0.38 & 2004ApJS...152..251 (INDO-US)\\
HD100407 & G7   & III   & 5010  & 2.93 & -0.04 & 1997A\&AS...124..299C (Cayrel)\\
HD100920 & G8.5 & III   & 4800  & 2.93 & -0.34 & 2004ApJS...152..251 (INDO-US) \\
HD101501 & G8   & V     & 5360  & 4.35 & -0.39 & 2004ApJS...151..387 (Ivanov)\\
HD102647 & A3   & V     & 8720  &      &       & 1995A\&AS...110..553 (Sokolov) \\
HD105707 & K2   & III   & 4320  & 2.16 & -0.13 & 1997A\&AS...124..299C (Cayrel) \\
HD106625 & B8   & III   &       &      &       &        \\
HD107259 & A2   & IV    & 9333  & 3.00 & 0.11  & 2004ApJS...152..251 (INDO-US)\\
HD107328 & K0   & IIIb  & 4380  & 2.39 & -0.48 & 2004ApJS...152..251 (INDO-US)\\
HD108767 & B9.5 & V     & 10350 &      &       & 1995A\&AS...110..553 (Sokolov) \\
HD109358 & G0   & V     & 5903  & 4.42 & -0.12 & 2004ApJS...151..387 (Ivanov)\\
HD109379 & G5   & II    & 5170  & 2.10 & -0.11 & 1997A\&AS...124..299C (Cayrel)\\
HD109387 & B6   & IIIpe &     &      &       &      \\
HD110379 & F0   &  V    & 7099  & 4.00 & -0.57 & 1997A\&AS...124..299C (Cayrel)\\
HD111812 & G0   & IIIp  &       &      & 0.01  & 2004ApJS...152..251 (INDO-US)\\
HD113139 & F2   & V     &       &      &       &   \\
HD113226 & G8   & III   & 4994  & 2.10 & 0.12  & 2004ApJS...151..387 (Ivanov)\\
HD113847 & K1   & III   & 4510  & 2.20 & -0.09 & 2004ApJS...152..251 (INDO-US)\\
HD113996 & K5   & III   & 3970  & 1.69 & -0.26 & 2004ApJS...151..387 (Ivanov)\\
HD114330 & A1   & Vs+.. & 9509  & 3.60 & -0.02 & 2004ApJS...152..251 (INDO-US)\\
HD114961 & M7   & III   & 3014  & 0.40 & -0.81 & 2004ApJS...152..251 (INDO-US)\\
HD115604 & F3   & III   & 7200  & 3.00 &  0.18 & 2004ApJS...152..251 (INDO-US)\\
HD115892 & A2   & V     & 9030  &      &       & 1995A\&AS...110..553 (Sokolov) \\
HD116656 & A2   & V     & 5793  &      &       & 2004ApJS...152..251 (INDO-US)\\
HD116658 & B1   & III   &       &      &       &        \\
HD116870 & K5   & III   &       &      &       &        \\
HD120315 & B3   & V     & 17200 &      &       & 1995A\&AS...110..553 (Sokolov) \\
HD120323 & M4.5 & III   &       &      &       &        \\
HD123139 & K0   & IIIb  & 4980  & 2.75 & 0.03  & 1997A\&AS...124..299C (Cayrel)\\
HD123299 & A0   & III   & 10080 & 3.30 & -0.56 & 2004ApJS...152..251 (INDO-US)\\
HD123657 & M4.5 & III   & 3452  & 0.90 & -0.03 & 2004ApJS...152..251 (INDO-US)\\
HD123934 & M2   & IIIa  &       &      &       &        \\
HD124294 & K2.5 & IIIb  & 4120  & 2.06 & -0.39 & 1997A\&AS...124..299C (Cayrel)\\
HD126661 & F0m  &       & 7754  & 3.50 & 0.10  & 2004ApJS...152..251 (INDO-US)   \\
\hline
\end{tabular}
\end{table}

\begin{table}
\caption{Physical Parameters of program stars -- Contd.}
\vspace{0.3cm}
\begin{tabular}{ccccccl} \hline
HD & Spectral & Luminosity &
T$_{eff}$($^{\circ}$K)& log$_{10}$(g) & (Fe/H) &Reference\\
 & Type & Class &
& & & \\
(1) & (2)& (3) & (4) & (5) & (6) & (7) \\ \hline

HD127665 & K3   & III   & 4260  & 2.22 & -0.17 & 2004ApJS...152..251 (INDO-US) \\
HD129116 & B3   & V     &       &      &       &        \\
HD129502 & F2   & III   & 6820  &      &       & 1995A\&AS...110..553 (Sokolov) \\
HD130025 & K0   &       & 5140  & 3.00 & -0.19 & 2004ApJS...152..251 (INDO-US) \\
HD130819 & F3   & V     & 6632  & 4.16 & -0.51 & 1997A\&AS...124..299C (Cayrel)\\
HD130841 & A3   & IV    &       &      &       &        \\
HD131156 & G7   & Ve    & 5500  & 4.60 & -0.15 & 2004ApJS...152..251 (INDO-US) \\
HD134083 & F5   & V     & 6632  & 4.50 & 0.32  & 2004ApJS...152..251 (INDO-US) \\
HD135722 & G8   & III   & 4834  & 2.45 & -0.39 & 2004ApJS...151..387 (Ivanov)\\
HD135742 & B8   & V     & 13250 &      &       & 1995A\&AS...110..553 (Sokolov) \\
HD136512 & K0   & III   & 4730  & 2.75 & -0.44 & 2004ApJS...152..251 (INDO-US) \\
HD141004 & G6   & V     & 5937  & 4.21 & -0.04 & 2004ApJS...152..251 (INDO-US) \\
HD141714 & G5   & III   & 5230  & 3.15 & -0.32 & 2004ApJS...152..251 (INDO-US) \\
HD141850 & M7   & III   &       &      &       &        \\
HD145328 & K1   & III   & 4678  & 2.50 & -0.14 & 2004ApJS...151..387 (Ivanov)\\
HD147165 & B1   & III   &       &      &       &        \\
HD148513 & K4   & III   & 4046  & 1.00 & 0.25  & 2004ApJS...151..387 (Ivanov) \\
HD161239 & G2   & IIIb  &       &      &       &        \\ \hline
\end{tabular}
\end{table}

    Fig. 6, shows spectra of six supergiant stars, covering a large range of
MK spectral type, and thus illustrates the basic dependence of spectral features. Fig 7
shows spectra of five giant stars again covering different spectral types. Similarly Fig 8
shows a series of six dwarf stars. All of these plots illustrate the change in basic features
with the temperature, gravity and metallicity.
We also attempt to show the quality of spectra by comparing some selected spectra with
the already published K band library by Wallace et al. (1997).

\begin{figure}
\epsfig{file=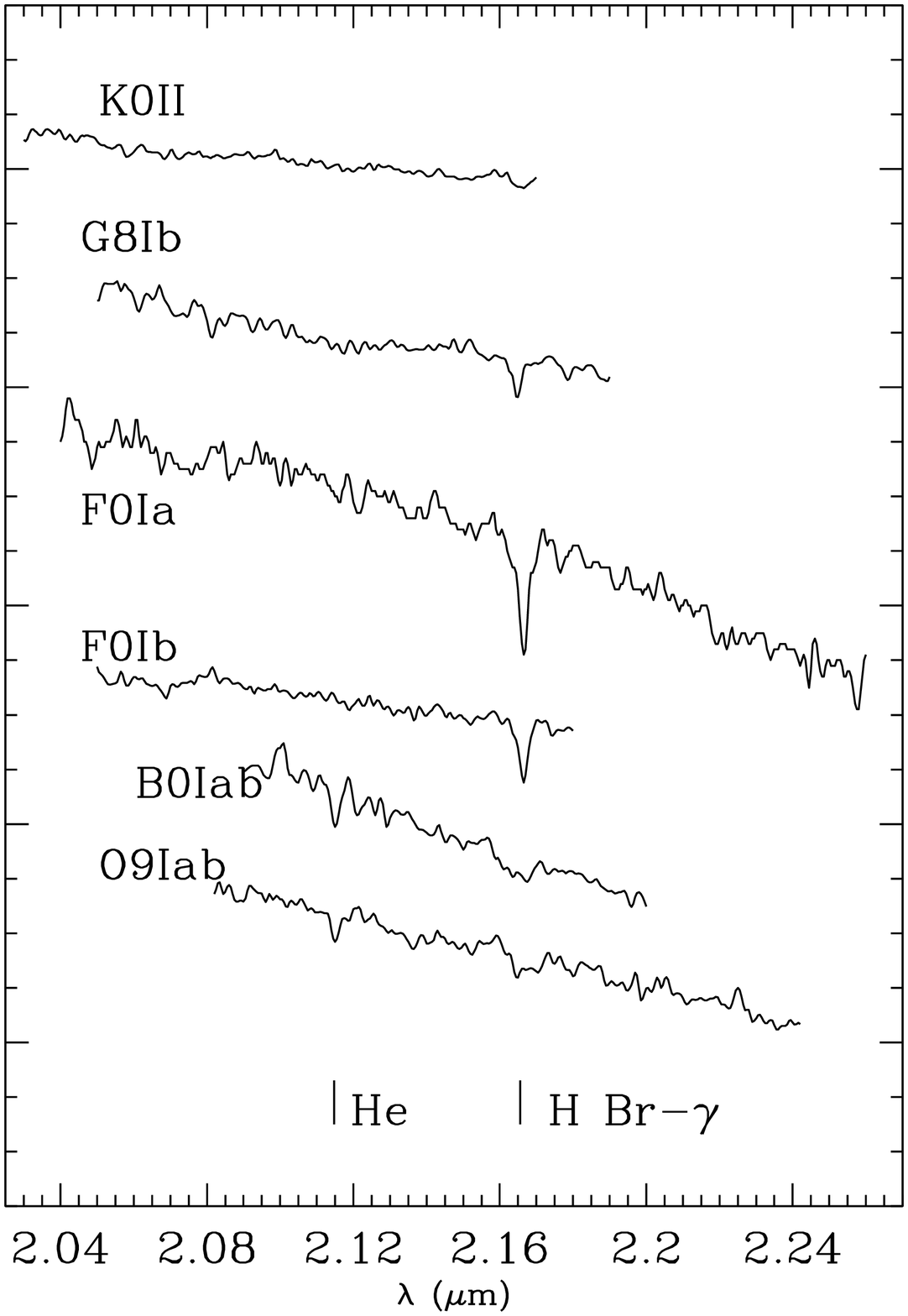,height=18cm,width=15cm}
\caption{Spectra of six supergiant stars, covering a large range of
MK spectral type, are plotted to illustrate the basic dependence
of spectral features. The stars plotted, top to bottom, are
HR4255, HR2473, HR382, HR1865, HR1903 and HD1948. The spectral
types are listed on the spectra.}
\end{figure}

\begin{figure}
\epsfig{file=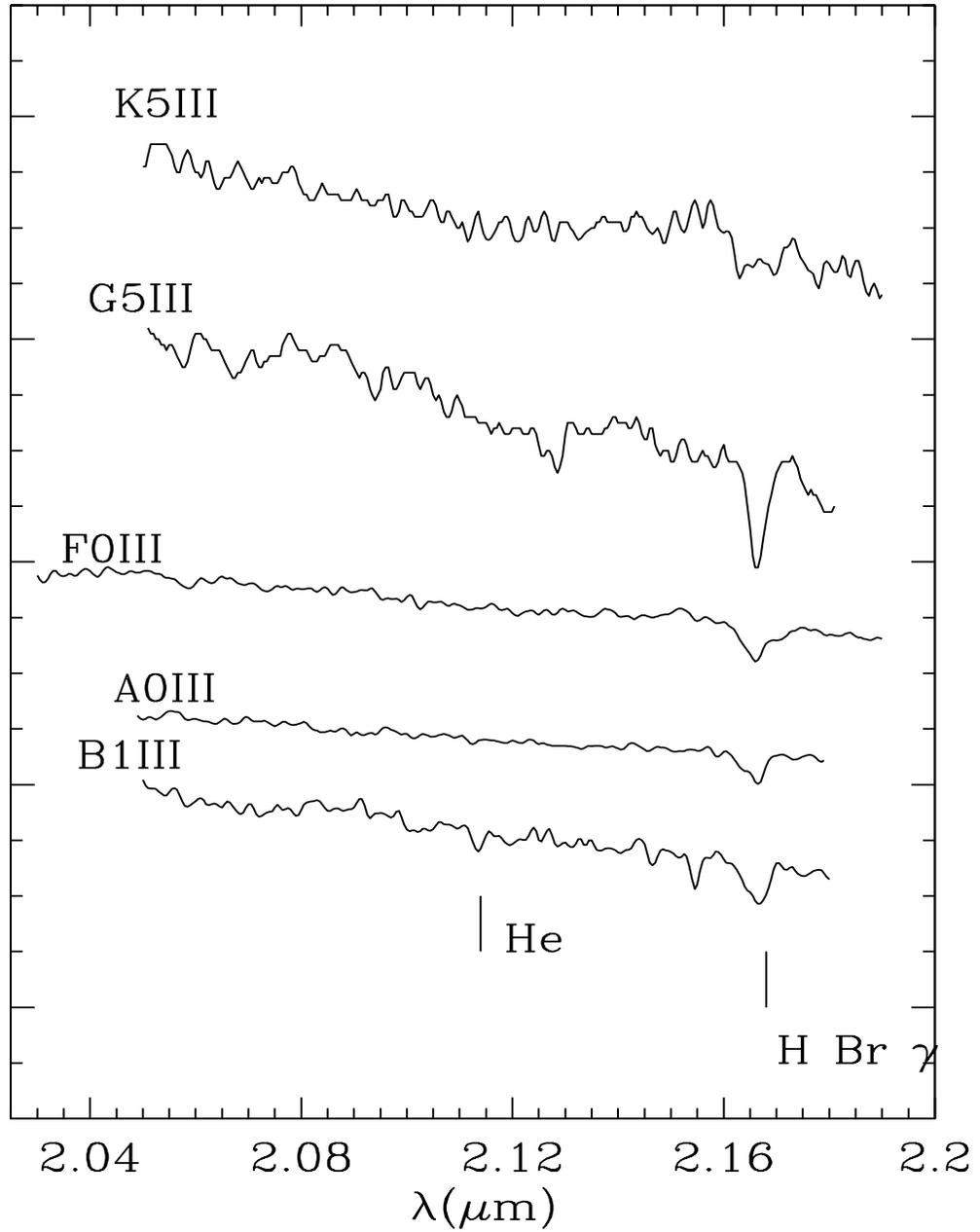,height=18cm,width=15cm}
\caption{Spectra of five giant stars, covering a large range of
MK spectral type, are plotted to illustrate the basic dependence
of spectral features. The stars plotted, top to bottom, are
HR4954, HR3323, HR4031, HR5291 and HR5056. The spectral
types are listed on the spectra.}
\end{figure}

\begin{figure}
\epsfig{file=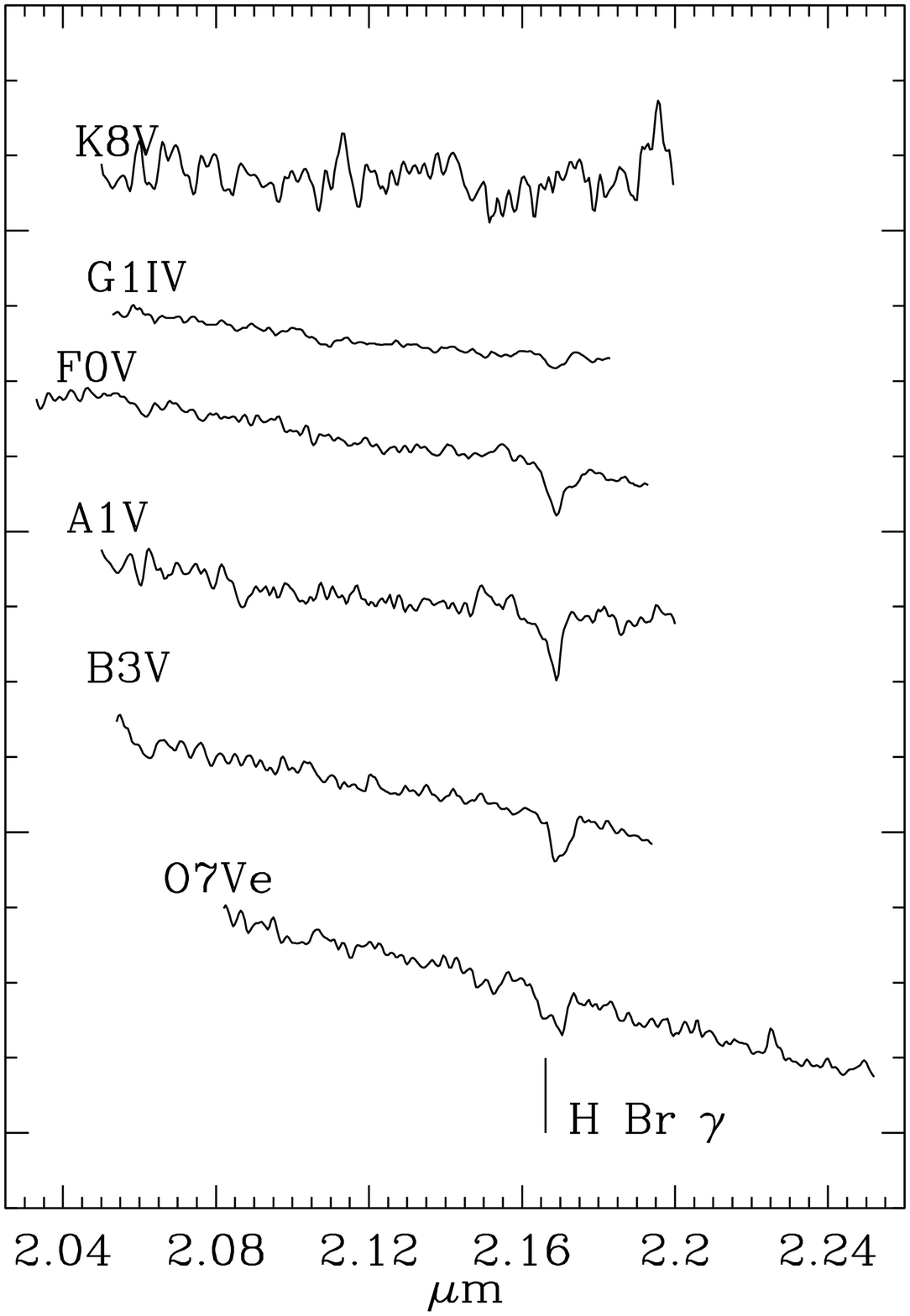,height=18cm,width=15cm}
\caption{Spectra of six dwarf stars, covering a large range of
MK spectral type, are plotted to illustrate the basic dependence
of spectral features. The stars plotted, top to bottom, are
HD88230, HR3176, HR4090, HR4963, HR5191 and HD2456. The spectral
types are listed on the spectra.}
\end{figure}

Following paragraphs describe the procedure that we have followed for
comparing the GIRT and Wallace data.

The block diagram in Fig. 9 depicts the steps carried out
on both libraries.
There are two steps to be performed on the Wallace et al. (1997).

(i) conversion of wavenumber vs. relative flux to wavelength vs. relative flux

(iii) fitting a continuum to respective T$_{eff}$ of each star and
spline fitting for binning at 5 \AA~ steps

\begin{figure}
\epsfig{file=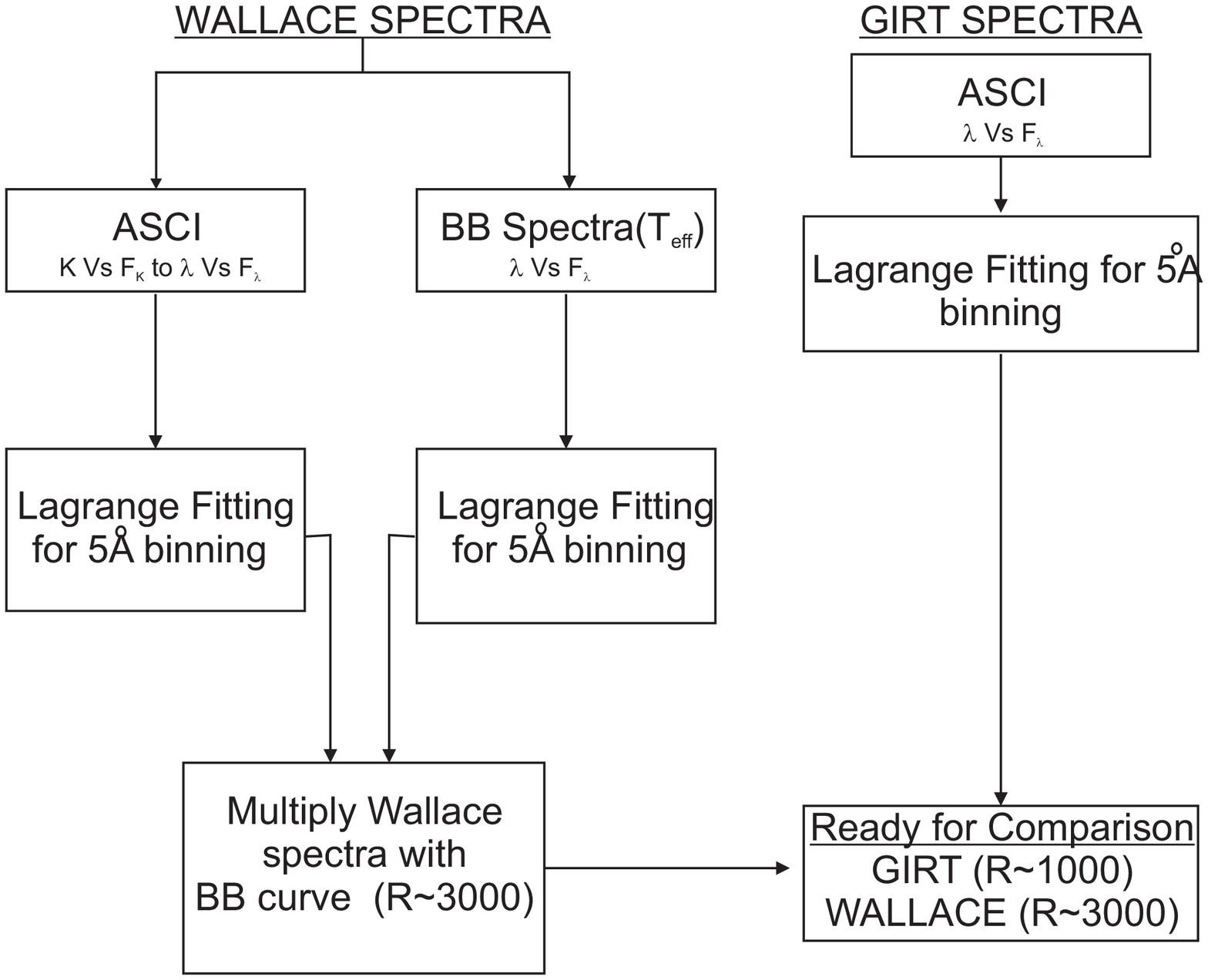,height=18cm,width=15cm}
\caption{Block diagram illustrating the steps involved in comparison of
GIRT and Wallace et al. (1997) libraries}
\end{figure}

These steps were performed by writing a common algorithm which could
run uniformly on the Wallace et al. (1997) library.
The T$_{eff}$ values were taken from 'Astronomical Hand Book' by K. R. Lang
and the black body spectra was generated by IRAF {\it{mk1dspec}} task
for K band region.

Fig. 10 shows a sample of some of the common stars in GIRT and
Wallace et al. library with good matching of the spectral features
as evident from the correlation coefficient {\it{r}} values. This
plot covers most of the main spectral types. It may be noted that
the resolution of both the spectra is not same viz. GIRT $\sim$ 1000
and Wallace $\sim$ 3000.

\begin{figure}
\epsfig{file=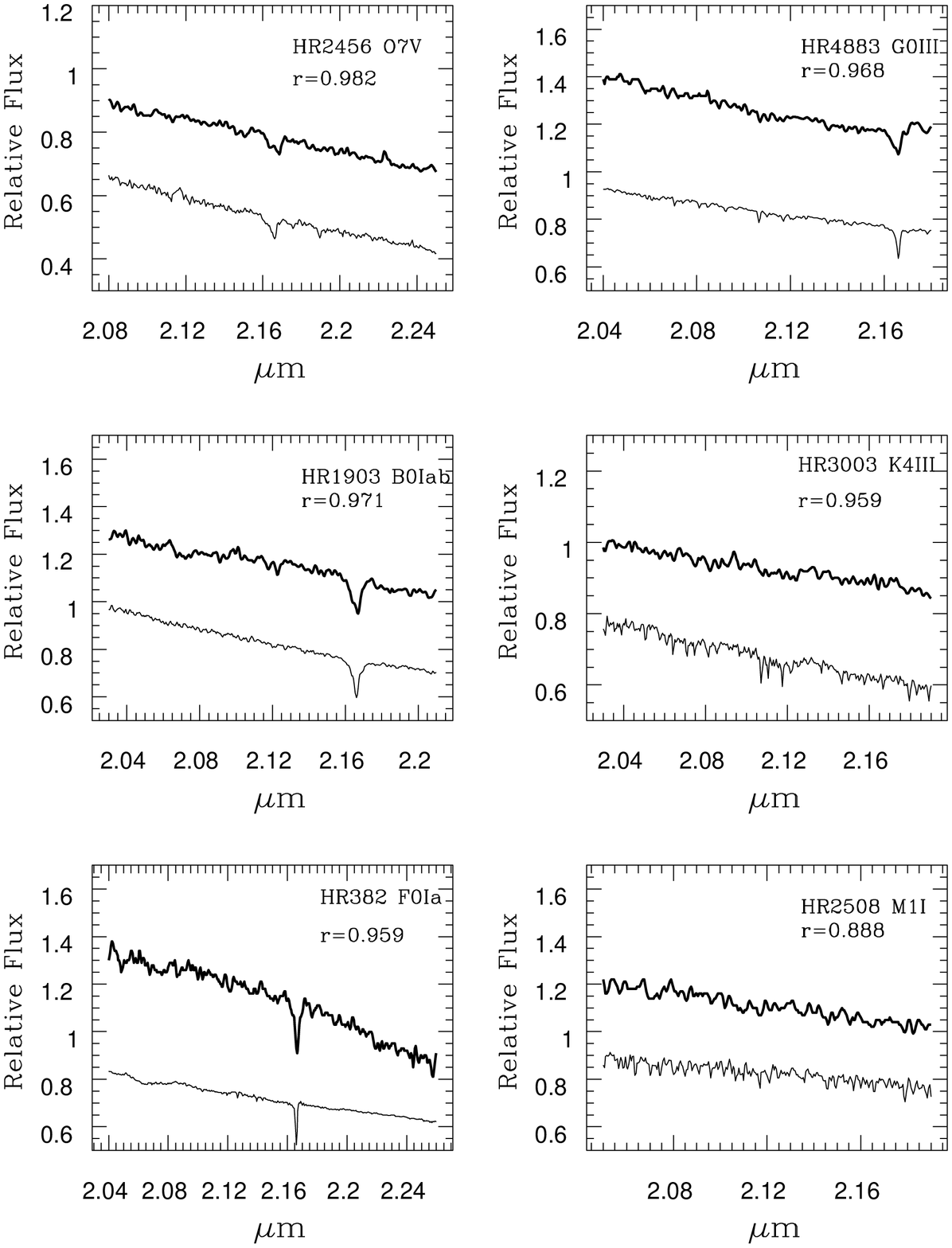,height=18cm,width=15cm}
\caption{Spectra of a selection of common spectra from Wallace et al. 1997
({\it thin} lines) and GIRT ({\it thick} lines) libraries. Please note
that the two spectra in each panel have been offset purposely for
sake of clarity and the flux values are relative.}
\end{figure}

In conclusion, we may mention that this library of 114 stellar
spectra in the NIR K band has been carefully checked for its
consistency with earlier published libraries and provides a larger
database with extended spectro-luminosity coverage for usage in
stellar population synthesis work and other applications as well as
complimenting large optical libraries. However it is very important
to have large spectral range to cover the entire spectra in K band
region, we would like to mention that we will carry out similar
observations to overcome spectral feature identification problem
either through using lamp spectra or choosing the right time of the year
when OH lines can be registered in sky frame.

\section*{Acknowledgements}
The research was partly funded by a grant from ISRO RESPOND. The research work
at the Physical Research Laboratory is funded by the Department of Space, Government
of India. This paper has made use of the SIMBAD database, operated at CDS, Strasbourg,
France.


\begin{thebibliography}{10}

\bibitem{} Blum, R. D., Ramond, T. M., Conti, P. S., Figer, D. F., Sellgren, K. 1997,
\newblock {\it AJ}, {\bf 113}, 1855.
\bibitem{} Bouchet, P., Manfroid, J., Schmider, F. -X., 1991,
\newblock {\it A\&AS}, {\bf 91}, 409.
\bibitem{} Bruzual, G. \& Charlot, S. 2003,
\newblock {\it MNRAS}, {\bf 344}, 1000.
\bibitem{} Carter, B. S. 1990, MNRAS, 242, 1.
\bibitem{} D. Thomas \& C. Maraston, 2003,
\newblock {\it A \& A}, {\bf 401}, 429.
\bibitem{} E. Ma\'rmol-Queralto\', N. Cardiel, et al. 2005,
\newblock {\it Serie de Conferencias}, {\bf 24}, 258.
\bibitem{} Fluks, M. A., Plez, B., The, P. S., Winter, D. de, Westerlund, B. E.,
Steenman, H. C. 1994,
\newblock {\it A\&AS}, {\bf 105}, 311.
\bibitem{} Ivanov, V. D., Reike, M. J., Englebracht, C. W., Alonso-Herrero, A.,
Reike, G. H., Luhman, K. L. 2004,
\newblock {\it ApJS}, {\bf 151}, 397.
\bibitem{} Johnson, H . J., \& Me\'ndez, M. E. 1970,
\newblock {\it AJ}, {\bf 75},785.
\bibitem{} Jones, L. A., 1999, Phd Thesis, University of North Carolina.
\bibitem{} Kleinman, S. G., \& Hall, D. N. B. 1986,
\newblock {\it ApJS}, {\bf 62},501.
\bibitem{} Koornneef, J. 1983,
\newblock {\it A\&AS}, {\bf 51}, 489.
\bibitem{} Lan\c{c}on, A., Rocca-Volmerange, B. 1992,
\newblock {\it A\&AS}, {\bf 96}, 593.
\bibitem{} Lang, K. R. 1992,
\newblock Astrophysical Data: Planets \& Stars, Springer-Verlag, New York.
\bibitem{} Le Borgne, D., Rocca-Volmerange, B., et al. 2004,
\newblock {\it A\&A}, {\bf 425}, 881.
\bibitem{} Le Borgne, J. F., Bruzual, G., Pello, R. et al. 2003,
\newblock {\it A\&A}, {\bf 402}, 433.
\bibitem{} Meyer, M. R., Edwards, S., Hinkle, K. H., Strom, S. E. 1998,
\newblock {\it ApJ}, {\bf 508}, 397.
\bibitem{} Prugniel, Ph., Soubiran, C. 2001,
\newblock {\it A\&A}, {\bf 369}, 1048.
\bibitem{} Prugniel, Ph., Soubiran, C. , Koleva, M., Le Borgne, D. 2007,
\newblock {\it Astro-ph}, 3658.
\bibitem{} Ranade, A., Gupta R., Ashok N. M., Singh H. P. 2004,
\newblock {\it BASI}, {\bf 32}, 311 (Paper I).
\bibitem{} Sa\'nchez-Bla\'zquez, P., Gorgas, J. et al. 2006,
\newblock {\it A\&A}, {\bf 457}, 809.
\bibitem{} Valdes, F., Gupta, R., Rose, J. A., Singh, H. P., Bell, D. J. 2004,
\newblock {\it ApJS}, {\bf 152}, 251.
\bibitem{} A. Vazdekis and N. Arimoto 1999,
\newblock {\it ApJ}, {\bf 525}, 144.
\bibitem{} Wallace, L., Hinkle K. 1997,
\newblock {\it ApJS}, {\bf 111}, 445.
\bibitem{} Wallace, L., Meyer, M. R., Hinkle, K., Edwards, S. 2000,
\newblock {\it ApJ}, {\bf 535}, 325.
\end{thebibliography}
\end{document}